\documentclass[journal,12pt,onecolumn,draftclsnofoot]{IEEEtran}
\usepackage[T1]{fontenc}
\usepackage{times}
\usepackage{color}
\usepackage{amsmath}
\usepackage{amssymb}
\usepackage{verbatim}
\usepackage{stmaryrd}
\usepackage{stackrel}
\usepackage{graphicx}
\usepackage{cite}
\usepackage{cases}
\usepackage{caption}
\usepackage{stackrel}
\usepackage{multicol}
\usepackage[USenglish]{babel}
\usepackage{array}
\usepackage{multirow}
\usepackage{amsfonts}
\usepackage{float}
\usepackage{balance}
\usepackage{ragged2e}
\usepackage{subfigure}
\usepackage{physics}
\usepackage[ruled,linesnumbered]{algorithm2e}
\usepackage{stfloats}
\usepackage{tikz}
\usepackage{autobreak}
\usepackage{gensymb}
\usepackage{booktabs}
\SetAlCapNameFnt{\scriptsize}
\SetAlCapFnt{\scriptsize}
\renewcommand{\raggedright}{\leftskip=0pt \rightskip=0pt plus 0cm}
\raggedright

\newtheorem{remrk}{\textbf{\textit{Remark}}}

\newtheorem{cor}{\textbf{\textit{Corollary}}}
\newtheorem{prop}{\textbf{\textit{Proposition}}}
\allowdisplaybreaks

\begin{document}
\title{Path Planning for Cellular-Connected UAV: A DRL Solution with Quantum-Inspired Experience Replay}
\author{Yuanjian~Li, A. Hamid~Aghvami~\IEEEmembership{Fellow,~IEEE} and Daoyi Dong~\IEEEmembership{Senior Member,~IEEE}\thanks{Yuanjian Li and A. Hamid Aghvami are with Centre for Telecommunications Research (CTR), King's College London, London WC2R 2LS, U.K. (e-mail: yuanjian.li@kcl.ac.uk; hamid.aghvami@kcl.ac.uk).}
\thanks{Daoyi Dong is with the School of Engineering and Information Technology,
University of New South Wales, Canberra ACT 2600, Australia. (e-mail:
d.dong@adfa.edu.au).}
\thanks{This work has been submitted to the IEEE for possible publication. Copyright may be transferred without notice, after which
	this version may no longer be accessible.}}

\maketitle
\vspace{-.5cm}
\begin{abstract}
In cellular-connected unmanned aerial vehicle (UAV) network, a minimization problem on the weighted sum of time cost and expected outage duration is considered. Taking advantage of UAV's adjustable mobility, an intelligent UAV navigation approach is formulated to achieve the aforementioned optimization goal. Specifically, after mapping the navigation task into a Markov decision process (MDP), a deep reinforcement learning (DRL) solution with novel quantum-inspired experience replay (QiER) framework is proposed to help the UAV find the optimal flying direction within each time slot, and thus the designed trajectory towards the destination can be generated. Via relating experienced transition's importance to its associated quantum bit (qubit) and applying Grover iteration based amplitude amplification technique, the proposed DRL-QiER solution commits a better trade-off between sampling priority and diversity. Compared to several representative baselines, the effectiveness and supremacy of the proposed DRL-QiER solution are demonstrated and validated in numerical results. 
\end{abstract}
\vspace{-.2cm}
\begin{IEEEkeywords}
	\vspace{-.4cm}
	Drone, trajectory design, deep reinforcement learning, quantum-inspired experience replay.
\end{IEEEkeywords}
%\vspace{-.5cm}
\section{Introduction}
\IEEEPARstart{W}ith flexible mobility, low
cost and on-demand deployment, unmanned aerial vehicles (UAVs) have been widely
used in civilian scenarios, e.g., building safety inspections, disaster management, material transport and aerial photography \cite{xiao2019unmanned,mei2019cellular,liu2019multi}. 
%To better unleash the potentials of UAV-aided applications, it is of significance to further enhance wireless transmission qualities between UAVs and ground transceivers. On one hand, UAVs may have to receive control and command (C\&C) messages from the ground entities to achieve safe operations. On the other hand, UAVs may need to unload mission-oriented data (e.g, images and videos) to the ground receivers. 
In practice, simple point-to-point (P2P) wireless links over unlicensed spectrum are commonly utilized to support the communications between UAVs and ground nodes, leading to constrained communication performance \cite{liu2019multi}. To further enhance wireless transmission qualities between UAVs and ground transceivers, cellular-connected UAV technique is deemed as a promising solution, via adopting widely-existing terrestrial base stations (BSs) to help establish high-quality ground-to-air (G2A) transmission links \cite{zeng2021simultaneous,zeng2019path,zhang2018cellular}. With the help of today's mature cellular networks and authentication mechanisms, cellular-connected UAV can help achieve better reliability, security and transmission throughput for G2A communications. Besides, cellular-connected UAV solution is significantly cost-effective because no dedicated infrastructures for supporting G2A communications are needed to construct and worldwide cellular BSs can be reused to aid G2A transmissions.
%Most importantly, cellular-aided localization can offer a new degree of freedom to help achieve more robust UAV navigation performance, as a supplement to global positioning system (GPS).

Current cellular networks are genuinely established for serving user equipments on the ground, via downtilting the main lobe of BS's antenna towards the earth \cite{zeng2021simultaneous}. This characteristic can enhance cellular coverage facing the ground but the quality of cellular-aided G2A transmissions cannot be guaranteed in general. More severe inter-cell interferences (ICIs) introduced by line-of-sight (LoS)-dominated G2A links can further deteriorate the aerial coverage issue, compared to terrestrial communication scenario where non line-of-sight (NLoS) channels are most likely experienced \cite{mei2019cooperative}. Fortunately, the controllable mobility feature of UAV makes it possible to tackle the aforementioned aerial coverage obstacles via UAV trajectory planning, either by on-board algorithms or remote pilots. The UAV navigation approach takes advantage of an extra degree of freedom, i.e., UAV's mobility, to realize aerial coverage enhancement and thus poses less or even no requirements on reconstruction of existing cellular infrastructures. Zhang \textit{et al.} \cite{zhang2018cellular} studied cellular-connected UAV's mission completion time minimization problem via invoking graph theory and convex optimization to design the optimal flying trajectory from an initial location to a destination, subject to connectivity constraint of the G2A link. Zhan \textit{et al.} \cite{zhan2020energy} maximized data uploading throughput for cellular-connected UAV under constraints of energy cost and minimum transmission rate threshold, via path planning with the help of successive convex approximation (SCA) technique. Bulut \textit{et al.} \cite{bulut2018trajectory} proposed a dynamic programming solution to help cellular-connected UAV find the best travelling path, subject to a continuous disconnection duration restriction. However, standard off-line optimization approaches solving trajectory design problem suffer from inefficiency due to non-convex nature of the formulated optimization objective and the corresponding constraints, even under impractical assumptions where perfect knowledge of wireless environment is available, e.g., G2A channel model and BS antenna model. Fortunately, reinforcement learning (RL) serves as a good complement to traditional off-line optimization solutions, which is famous for the favourable ability of learning unknown environment in a trial-and-error manner. Up to date, RL-related techniques have been widely applied to help solve performance optimization problems for UAV-mounted networks, e.g., radio resource allocation, interference mitigation and path planning. Cui \textit{et al.} \cite{cui2019multi} investigated a real-time design on resource allocation for multiple-UAV network, in which multi-agent reinforcement learning (MARL) framework was proposed to realize optimal user selection, power allocation and sub-channel association. Zeng \textit{et al.}  \cite{zeng2021simultaneous} investigated an optimal UAV trajectory planning problem on minimizing the weighted sum of mission completion time and expected transmission outage duration, via deep reinforcement learning (DRL)-aided approaches.

Meanwhile, quantum theory has been proven to pose a positive impact on improving learning efficiency for artificial intelligence algorithms in general, and RL-related approaches in particular. Dong \textit{et al.} \cite{dong2008quantum} combined quantum parallelism into conventional RL frameworks (termed as quantum RL (QRL)), in which higher learning efficiency and better trade-off between exploration and exploitation were showcased. Furthermore, Dong \textit{et al.} \cite{dong2010robust} proposed quantum-inspired reinforcement learning (QiRL) to solve intelligent navigation problem for autonomous mobile robots, where probabilistic action selection method and novel reinforcement approach inspired by quantum phenomenon were integrated into standard RL frameworks. Paparo \textit{et al.} \cite{paparo2014quantum} showed that quadratic speed-up is achievable for intelligent agents, with the help of quantum mechanics. Dunjko \textit{et al.} \cite{dunjko2016quantum} extended traditional agent-environment framework into quantum region, while Saggio \textit{et al.} \cite{saggio2021experimental} demonstrated the first experimental result of QRL.
%Lamata \cite{lamata2017basic} conducted QRL on superconducting circuits with multiple quantum bits (qubits). Hu \textit{et al.} \cite{hu2019training} solved a representative RL problem, i.e., contextual multi-armed bandit, via training a quantum neural network with photonic quantum circuits, illustrating that QRL algorithms can be trained on quantum devices
In \cite{li2020quantum}, Li \textit{et al.} compared QRL with several RL frameworks in human decision-making scenarios, suggesting that value-based decision-making can be illustrated by QRL at both the behavioural and neural levels. In the field of wireless communications, Li \textit{et al.} \cite{li2021intelligent} investigated an optimal path planning problem for UAV-mounted networks, in which QiRL solution was demonstrated to offer better learning performance than conventional RL methods with $\epsilon$-greedy or Boltzmann action selection policy.

In this paper, we integrate several ideas in quantum mechanics and DRL techniques to solve intelligent trajectory planning problem for cellular-connected UAV networks. The main contributions of this paper are summarized as follows.

\begin{itemize}
	\item Different from the vast majority of existing literature, more practical G2A pathloss model based on one realization of local building distribution and directional antenna with fixed 3-dimensional (3D) radiation pattern are considered in this paper. Then, a cellular-connected UAV trajectory planning problem is formulated to minimize the weighted sum of flight time cost and the corresponding expected outage duration. Without prior knowledge of the wireless environment, the focused path planning problem is challenging to be tackled via conventional optimization techniques. Alternatively, the proposed optimization problem is mapped into Markov decision process (MDP) and solved by the proposed DRL solution with novel quantum-inspired experience replay (QiER).
	\item A novel QiER framework is coined to help the learning agent achieve better training performance, via a three-phase quantum-inspired process. Specifically, the quantum initialization phase allocates initial priority for the newly-recorded experiences, the quantum preparation phase generates the updated priority for the sampled transitions with the help of Grover iteration, and the quantum measurement phase outputs distribution of sampling probabilities to help accomplish the mini-batch training procedure.
	\item To demonstrate advantages offered by the proposed DRL-QiER solution, performance comparison against representative baselines is performed. Compared to DRL approach with standard experience relay (DRL-ER) or prioritized ER (DRL-PER), deep curriculum reinforcement learning (DCRL) method and simultaneous navigation and radio mapping (SNARM) strategy, simulation results demonstrate that the proposed DRL-QiER solution can achieve more efficient and steady learning performance. Nevertheless, the proposed DRL-QiER does not include extra neural networks like SNARM approach, and requires much less hyper-parameter tuning like DCRL or DRL-PER method, which means that it is easier and more robust for implementation. 
\end{itemize}

Although a similar cellular-connected UAV environment with \cite{zeng2021simultaneous} is considered, the main differences are: 1) UAVs' flying directions are enlarged from 4 to 8, which allows UAVs to travel more flexibly and thus better navigation performance may be achieved, but inevitably making the agent's neural networks more difficult to be trained; 
%2) prioritized experience replay (PER) technique is invoked to help the UAV learn the wireless environment more efficiently; 
and 2) an innovative DRL-QiER solution is proposed to solve the navigation task in a more efficient manner. Nevertheless, with the help of Grover iteration in quantum computation, we extend the QiER method in \cite{wei2021deep} from 2-dimensional (2D) discrete rotation to its 3D continuous alternative, which introduces fewer additional hyper-parameters and thus makes the QiER technique more flexible and reliable.

\textit{\textcolor{black}{Organization}}\textcolor{black}{: Section \uppercase\expandafter{\romannumeral2}
	presents the system model. Section \uppercase\expandafter{\romannumeral3}
	gives a brief overview on DRL. Section \uppercase\expandafter{\romannumeral4}
	briefly introduces quantum state and quantum amplitude amplification. Section
	\uppercase\expandafter{\romannumeral5} presents the proposed DRL-QiER solution. Simulation results
	are given in Section \uppercase\expandafter{\romannumeral6}, while
	conclusions are drawn in Section \uppercase\expandafter{\romannumeral7}.}
\section{System Model}
A downlink transmission scenario inside cellular-connected UAV network is considered, where a set $\mathcal{U}$ of $U$ UAVs is served by a set $\mathcal{B}$ of $B$ BSs within cellular coverage. These UAVs are supposed to reach a common destination from their respective initial locations, for accomplishing their own missions.\footnote{For example, one typical UAV application case is parcel collection. Various UAVs are launched from different costumers' properties carrying parcels to the local distribution centre of delivery firm. Besides, collision avoidance during UAVs' flights needs to be guaranteed, via separating UAV's operation spaces and keeping their flying altitudes higher than the tallest building.} Intuitively, each UAV should be navigated with a feasible trajectory, alongside which the corresponding time consumption should be the shortest and wireless transmission quality provided by the cellular network should be maintained satisfactorily. Without loss of generality, an arbitrary UAV (denoted as $u$ hereafter) out of these $U$ drones are concentrated for investigating the navigation task.\footnote{These UAVs share the same airspace and common location-dependent database, which means that the trained DRL model can be downloaded by the remaining UAVs, helping them accomplish their navigation tasks.} For clarity, the UAV's exploration environment is defined as a cubic subregion $\mathbb{A}:[x_{\text{lo}}, x_{\text{up}}]\times[y_{\text{lo}}, y_{\text{up}}]\times[z_{\text{lo}}, z_{\text{up}}]$, where the subscripts "lo" and "up" represent the lower and upper boundaries of this 3D airspace, respectively. Furthermore, the coordinate of the focused UAV at time $t$ should locate in the range of $\vec{q}_{\text{lo}}\preceq\vec{q}_{u}(t)\preceq\vec{q}_{\text{up}}$, where $\vec{q}_{\text{lo}}=(x_{\text{lo}}, y_{\text{lo}}, z_{\text{lo}})$, $\vec{q}_{\text{up}}=(x_{\text{up}}, y_{\text{up}}, z_{\text{up}})$ and $\preceq$ denotes the element-wise inequality. The initial location and the destination are given by $\vec{q}_{u}(I)\in\mathbb{R}^{1*3}$ and $\vec{q}_{u}(D)\in\mathbb{R}^{1*3}$, respectively. Therefore, the overall trajectory of this UAV's flight can be fully traced by $\vec{q}_{u}(t)=(x_{u}(t),y_{u}(t),z_{u}(t))$, starting from $\vec{q}_{u}(I)$ and ending at $\vec{q}_{u}(D)$. Besides, the location of arbitrary BS $b\in\mathcal{B}$ is indicated as $\vec{q}_{b}=(x_{b}, y_{b}, z_{b})$, where $\vec{q}_{\text{lo}}\preceq\vec{q}_{b}\preceq\vec{q}_{\text{up}}$. 
%Note that the locations of BSs are static, which will later be specified in the simulation section.

\subsection{Antenna Model}
%While G2A channel model is of importance for characterize the performance of G2A links, antenna model for cellular BSs is vital as well. 
Terrestrial transmission usually assumes that the distance between transceivers is much greater than the height difference of their antennas. In this regard, antenna modelling for terrestrial communications mainly concerns 2D antenna gain on the horizontal domain. Unfortunately, 2D antenna modelling is not sufficiently feasible for G2A transmissions, where high-altitude UAVs are involved. 
%More practically, 3D antenna gain should be considered for G2A transmissions, which takes both the azimuth and elevation angles into account.

\begin{figure}[htbp]
	\centering  
	\subfigcapskip = -.5cm
	\vspace{-.5cm}
	\subfigure[\tiny Coordinate system of ULA]{
		\label{ULA}
		\includegraphics[scale=0.6]{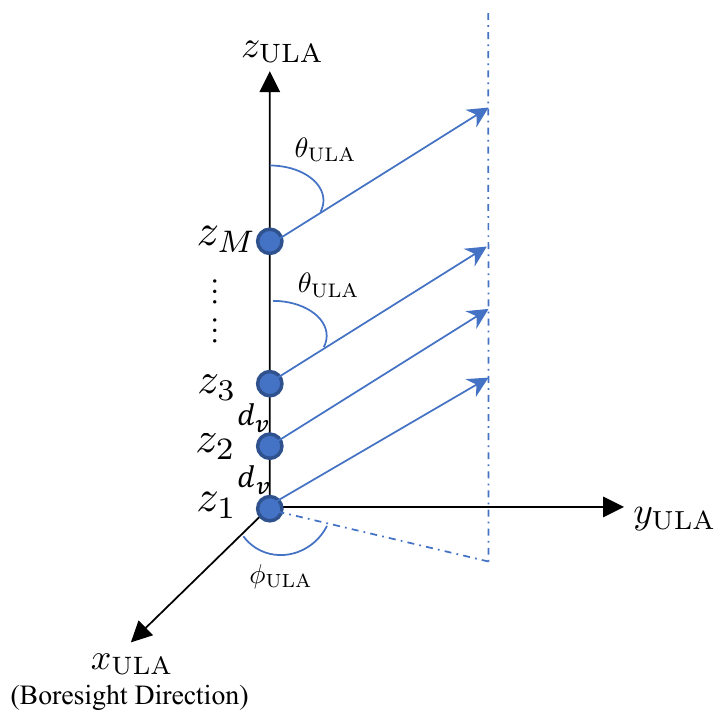}}
	\hspace{3cm}
	\subfigure[\tiny Vertical pattern at boresight]{
		\label{Vertical radiation pattern with azimuth angle}
		\includegraphics[scale=0.4]{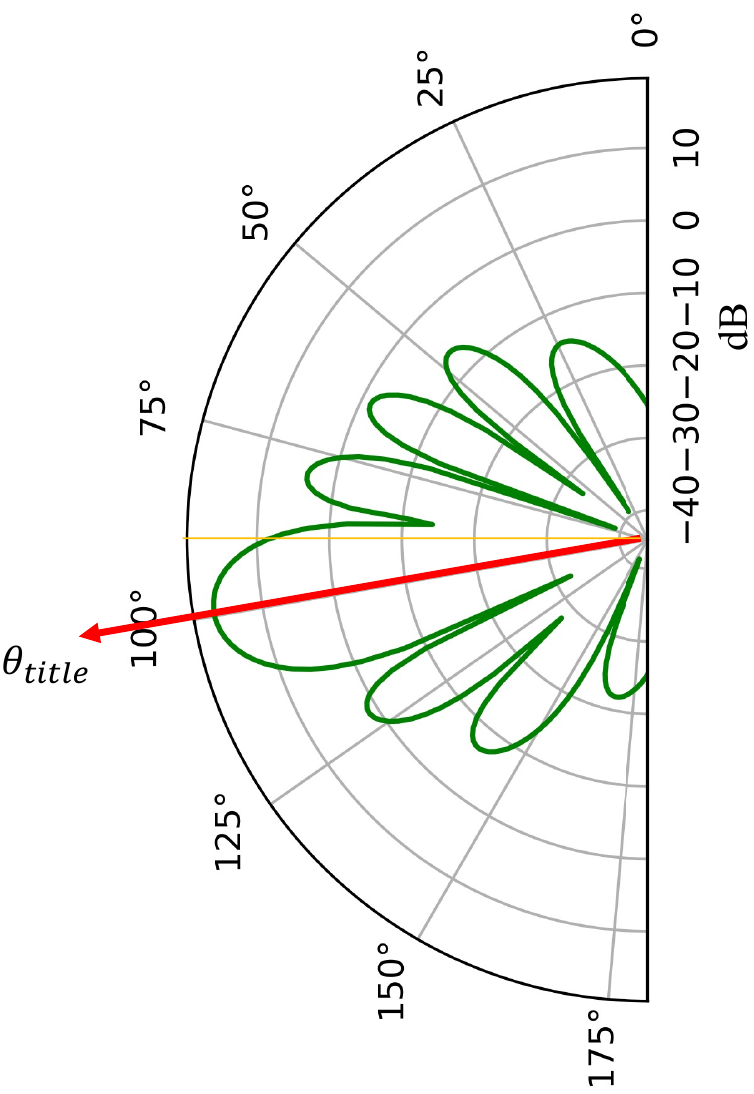}}
	\captionsetup{font={scriptsize}}
	\vspace{-.5cm}
	\caption{Demonstration of ULA's coordinate system and vertical radiation pattern}
	\label{Demonstration of vertical radiation pattern and sectorization}	
\end{figure}

In compliance with BS's antenna modelling of current cellular networks, directional antenna with fixed 3D radiation pattern is assumed to be equipped at each BS. Following standard sectorization, each BS is portioned to cover three sectors. Therefore, there are $3B$ sectors in total within the interested airspace $\mathbb{A}$. Specifically, it is assumed that three vertically-placed $M$-element uniform linear arrays (ULAs) are equipped by each BS with boresights directed to their corresponding sectors covered by this BS, subject to the 3GPP specification on cellular BS's antenna model \cite{ThreeGPPantenna2017}. 
%as $\left\{60\degree,180\degree,300\degree\right\}$
In individual and independent coordinate system of each ULA (e.g., Fig. \ref{ULA}), antenna element's placing location is denoted as $(0,0,z_{m})$, where $m=\{1,2,\dots,M\}$. 

Then, the wave factor of ULA can be given by 
\begin{equation}
\vec{k} = \frac{2\pi}{\lambda}\left(\sin\theta_{\text{ULA}}\cos\phi_{\text{ULA}},\sin\theta_{\text{ULA}}\sin\phi_{\text{ULA}},\cos\theta_{\text{ULA}}\right),
\end{equation}
where $\lambda=c/f_{c}$ represents the wavelength, $c$ denotes the light speed and $f_{c}$ indicates the carrier frequency. Furthermore, the steering vector can be derived as 
\begin{equation}
\vec{sv} = \left[\exp(-j\vec{k}(0,0,z_{1})^{T}),\dots,\exp(-j\vec{k}(0,0,z_{M})^{T})\right]^{T}.
\end{equation}
Suggested by 3GPP, vertical and horizontal radiation patterns in dB of each ULA are given by
\begin{equation}
A_{V}\left(\theta_{\text{ULA}},\phi_{\text{ULA}}=0\degree\right) = -\min\left\{12\left(\frac{\theta_{\text{ULA}} - 90\degree}{\Theta_{3\text{dB}}}\right)^{2}, 30\text{dB}\right\},
\end{equation}
\begin{equation}
A_{H}\left(\theta_{\text{ULA}}=90\degree,\phi_{\text{ULA}}\right) = -\min\left\{12\left(\frac{\phi_{\text{ULA}}}{\Phi_{3\text{dB}}}\right)^{2}, 30\text{dB}\right\},
\end{equation}
respectively. Then, each ULA's 3D element pattern in dB can be calculated as
\begin{equation}
A\left(\theta_{\text{ULA}},\phi_{\text{ULA}}\right) = -\min\left\{-\left[A_{V}\left(\theta_{\text{ULA}},\phi_{\text{ULA}}=0\degree\right)+A_{H}\left(\theta_{\text{ULA}}=90\degree,\phi_{\text{ULA}}\right)\right], 30\text{dB}\right\}. 
\end{equation}
Note that each antenna element of a ULA is directional, specified by half-power beamwidths $\Theta_{3\text{dB}}$ and $\Phi_{3\text{dB}}$ for the vertical and horizontal dimensions, respectively. To suppress ICIs in cellular networks, the main lobe of ULA's radiation pattern should be electrically steered by $\theta_{etilt}\in[0\degree,180\degree]$, where $\theta_{etilt}=90\degree$ means perpendicular to the ULA. To achieve the steering angle $\theta_{etilt}$, fixed phase shift for each antenna element of ULA can be executed, for which the complex coefficient of the $m$-th antenna element is given by
\begin{equation}
\omega_{m} = \frac{1}{M}\exp\left[-j\frac{2\pi}{\lambda}(m - 1)d_{v}\cos{\theta_{etilt}}\right],
\label{steering_weight}
\end{equation}
where $d_{v}$ indicates the vertical elements' spacing distance. Furthermore, the array factor can be formulated as 
\begin{equation}
AF = \sum_{m = 1}^{M}\omega_{m}\exp(-j\vec{k}(0,0,z_{m})^{T}) 
=\vec{\omega}\vec{sv},
\end{equation}
where $\vec{\omega}=(\omega_{1},\dots,\omega_{M})^{*}$ is the weight vector and the superscript $*$ indicates the complex conjugate. In the end, the 3D antenna gain of each ULA in dB can be calculated as 
\begin{equation}
G\left(\theta_{\text{ULA}},\phi_{\text{ULA}}\right) = 10\lg\left(\vert\sqrt{10^{\frac{A\left(\theta_{\text{ULA}},\phi_{\text{ULA}}\right)}{10}}}AF\vert^{2}\right).
\end{equation}

Fig. \ref{Vertical radiation pattern with azimuth angle} illustrates an example for $\theta_{etilt} = 100\degree$, under parameter setting $\Theta_{3\text{dB}} = \Phi_{3\text{dB}} = 65\degree$, $d_{v} = \lambda/2$ and $M = 8$. It is straightforward to observe that the main lobe is downtilted towards the ground for serving terrestrial communications, and the upper side lobes can be utilized to support G2A transmissions.
Denote $i\in\{1,\dots,3B\}$ as the label of sectors in the considered region. Then, the transmit antenna gain from arbitrary sector to the UAV can be explicitly determined by UAV's location, denoted as $G^{i}\left[\vec{q}_{u}(t)\right] = G\left(\theta_{iu},\phi_{iu}\right)$, where $\theta_{iu}$ and $\phi_{iu}$ can be obtained via taking $\vec{q}_{u}(t)$, the location of ULA for sector $i$ and the ULA's boresight direction into account.\footnote{Note that the location of ULA for sector $i$ is assumed to be the same as its associated BS, which is a reasonable consideration because the distance among ULAs on the BS is much smaller than that between the UAV and the BS.} 
\begin{comment}
Denote $c\in\{3(b-1)+1,3(b-1)+2,3(b-1)+3\}$ as the label of sector associated with BS $b\in\mathcal{B}$, where the total number of sectors equals to $3B$. Then, the transmit antenna gain from arbitrary sector of each BS to the UAV can be fully determined by UAV's current location, which is denoted as $G_{\vec{q}_{b}}^{c}\left[\vec{q}_{u}(t)\right] = G\left(\theta_{bu},\phi_{bu}\right)$, in which $\theta^{c}_{bu}$ and $\phi^{c}_{bu}$ can be obtained 
\end{comment}

\subsection{Pathloss Model}
\label{Sec_channnel_models}
Different from terrestrial transmissions, G2A links are more likely to experience LoS pathloss. In this subsection, the adopted G2A channel model will be interpreted.

According to 3GPP urban-macro (UMa) pathloss model \cite{ThreeGPP2017}, the G2A pathloss in dB from sector $i$ to the UAV at time $t$ is given by 
\begin{equation}
\text{PL}^{i}\left[\vec{q}_{u}(t)\right]=
\begin{cases}
28.0+22\log_{10}\left(d_{iu}\right)+20\log_{10}\left(f_{c}\right),&\text{if LoS} \\
-17.5+\left[46-7\log_{10}\left(z_{u}(t)\right)\right]
\log_{10}\left(d_{iu}\right)+20\log_{10}\left(\frac{40\pi f_{c}}{3}\right),&\text{if NLoS}
\end{cases},
\label{pathloss_LoS_or_NLoS}
\end{equation}
where $d_{iu}=\vert\vert\vec{q}_{u}(t)-\vec{q}_{i}\vert\vert_{2}$ outputs the Euclidean distance between the UAV and the location of ULA for sector $i$.
%$r_{iu}=\sqrt{d_{iu}^2-z_{u}(t)^2}$ and
%Nakagami-{$m$} fading\footnote{In contrast to terrestrial communication scenarios where Rayleigh fading is widely applied to model small-scale fading, Rician or Nakagami-$m$ fading is more suitable to track the characteristics of G2A small-scale fading since the LoS-dominated G2A channels.} is invoked to represent the small-scale fading component for G2A channels, denoted as $\vec{h}_{bu}(t)\in\mathbb{C}^{1\times M}$. Note that the shape factor $m$ varies with its corresponding type of large-scale fading, i.e., LoS or NLoS. 
%In this paper, we set $m=1$ (equivalent to Rayleigh fading) for $\text{PL}_{\text{NLoS}}$ and $m=3$ for $\text{PL}_{\text{LoS}}$.

\begin{figure}[htbp]
	\centering  
	\subfigcapskip = -.5cm
	\vspace{-.5cm}
	\subfigure[\tiny Local building distribution]{
		\label{level.sub.1}
		\includegraphics[scale=0.38]{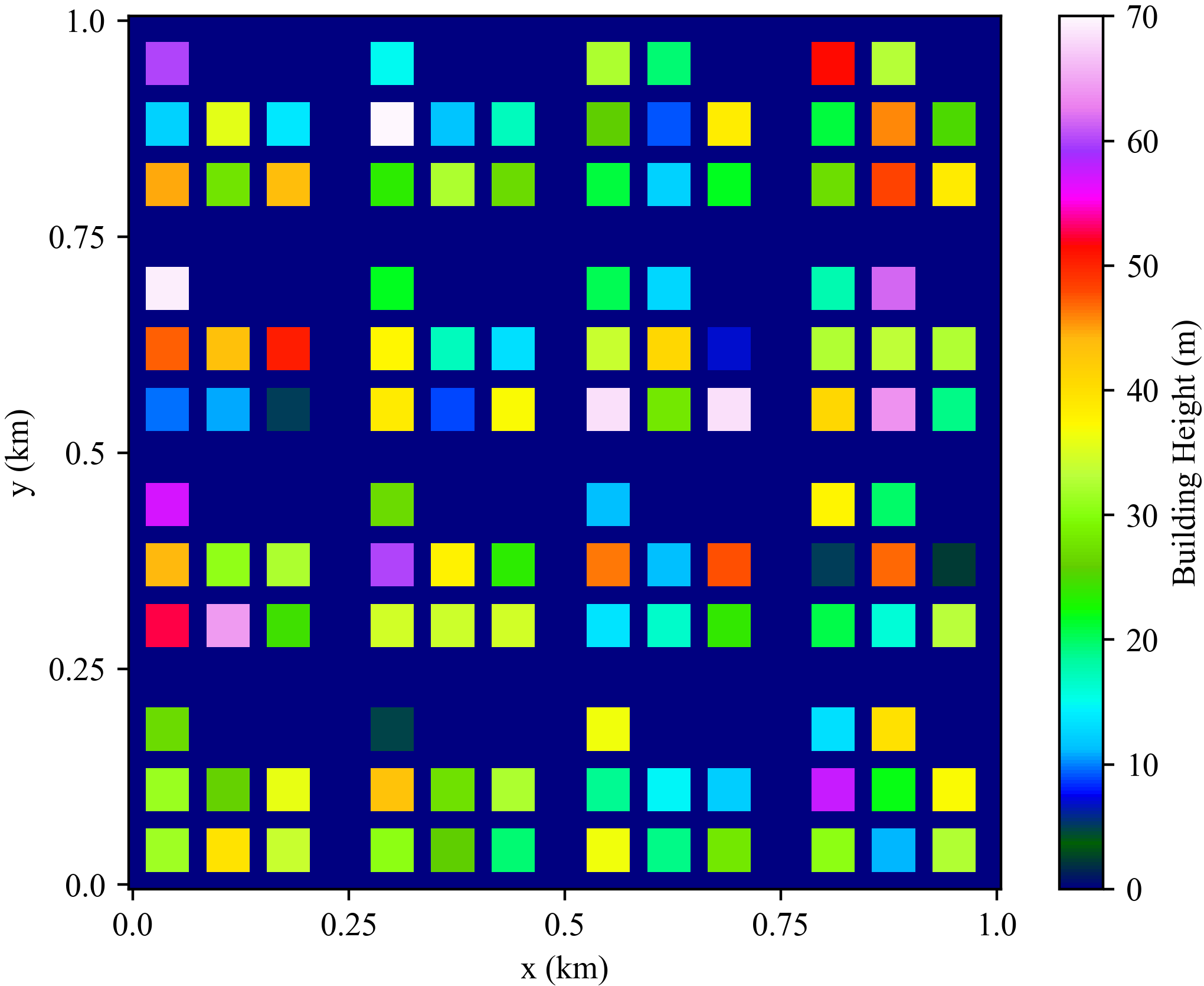}}\hspace{3cm}
	\subfigure[\tiny 3D view of local building distribution]{
		\label{level.sub.2}
		\includegraphics[scale=0.40]{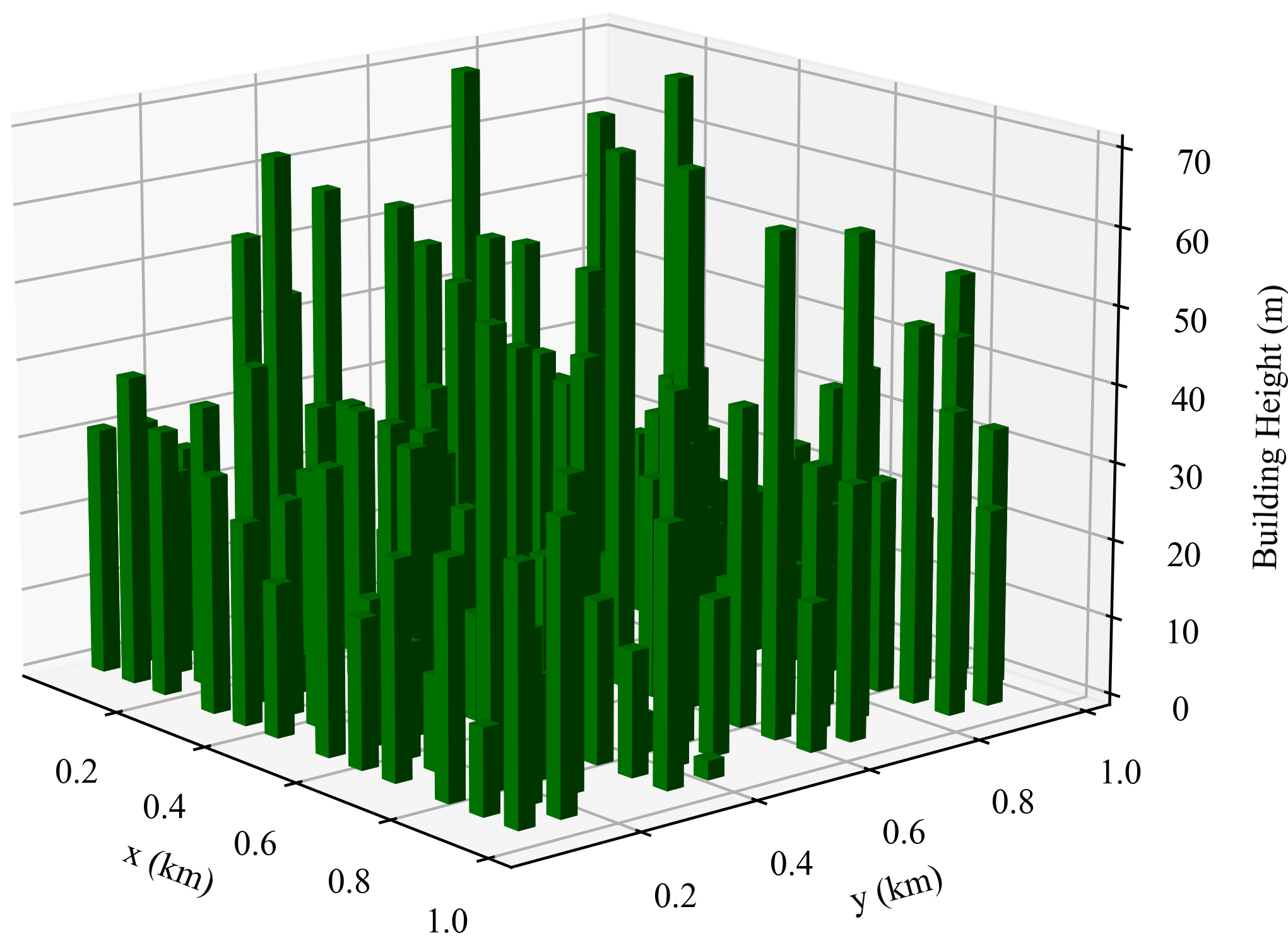}}
	\captionsetup{font={scriptsize}}
	\vspace{-.5cm}
	\caption{The building distribution under consideration}
	\vspace{-1cm}
	\label{Building Distribution}	
\end{figure}

To practically trace the type of G2A pathlosses, building distribution in the interested airspace $\mathbb{A}$ should be taken into consideration. Fig. \ref{Building Distribution} illustrates an example of local building distribution, including their horizontal locations on the ground and heights (Fig. \ref{level.sub.1}), as well as the corresponding 3D view (Fig. \ref{level.sub.2}). With given building distribution, the type of large-scale pathloss of G2A channels for UAV at arbitrary location $\vec{q}_{u}(t)$, i.e., LoS or NLoS in (\ref{pathloss_LoS_or_NLoS}), can be accurately determined via checking the potential blockages between the UAV and sectors.\footnote{Note that our method generating G2A pathloss is more practical than the widely-used probabilistic G2A channel model in current literature because the later can only characterize the average G2A pathloss rather than its real counterpart.} 

\subsection{SINR at UAV}
With the aforementioned antenna and pathloss models, the received signal of the focused UAV $u$ at arbitrary location $\vec{q}_{u}$ over time $t$ can be formulated as
\begin{equation}
y_{u}(t)\!=\sum_{i=1}^{3B}\sqrt{10^{\frac{G^{i}\left[\vec{q}_{u}(t)\right]-\text{PL}^{i}[\vec{q}_{u}(t)]}{10}}}h_{iu}x_{i}(t) + n_{u}(t),
\end{equation}
where $x_{i}(t)\sim\mathcal{CN}(0,P_{i})$ is the emitted message from sector $i$ to the UAV with average transmit power $P_{i}$, $h_{iu}$ represents the corresponding small-scale fading channel\footnote{This paper aims to develop a UAV navigation method for arbitrary small-scale channel model. Hence, we do not specify the type of small-scale fading here, e.g., Rayleigh \cite{zhang2008cooperative,dong2004optimal,ma2008blind,sheng2015energy,tang2014resource}, Rician \cite{cheng2013wideband,wen2011sum} or Nakagami-$m$ \cite{upadhyay2011performance}.} and $n_{u}(t)\sim\mathcal{CN}(0,\sigma^2)$ denotes the received additive complex Gaussian noise (AWGN) at the UAV. Note that the explicit type of pathloss, i.e., LoS or NLoS, can be determined via checking possible blockages according to one realization of local building distribution as mentioned in Section \ref{Sec_channnel_models}. Assume that the UAV is associated with sector $\hat{i}$ at time $t$, the instantaneous signal-to-interference-plus-noise ratio (SINR) at the UAV can be derived as 
\begin{equation}
\Gamma_{u}(t) = \frac{P_{\hat{i}}10^{\frac{G^{\hat{i}}\left[\vec{q}_{u}(t)\right]-\text{PL}^{\hat{i}}[\vec{q}_{u}(t)]}{10}}\vert h_{\hat{i}u}\vert^2}{I_{u}(t)+\sigma^2},\label{SINR_u}
\end{equation}
where $I_{u}(t)=\sum_{i\neq \hat{i}}P_{i}10^{\frac{G^{i}\left[\vec{q}_{u}(t)\right]-\text{PL}^{i}[\vec{q}_{u}(t)]}{10}}\vert h_{iu}\vert^2$ means the ICIs from un-associated sectors.\footnote{This paper focuses on the worst case where universal frequency reuse is assumed, which means that all the non-associated co-channel sectors will be taken into account as the sources of ICIs.}
%the transmission rate in bps/Hz at DUE $u$ for time slot $t$ can be expressed as
%\begin{equation}
%RT_{u}(t)=\log_{2}(1+\gamma_{u}(t))
%\end{equation}

\subsection{Problem Formulation}
The received SINR (\ref{SINR_u}) is a random variable because of the randomness introduced by small-scale fadings, with given UAV coordinate $\vec{q}_{u}(t)$ and cell association $\hat{i}(t)$. Therefore, the corresponding transmission outage probability (TOP) can be formulated as a function of $\vec{q}_{u}(t)$ and $\hat{i}(t)$, i.e., $TOP_{u}\{\vec{q}_{u}(t), \hat{i}(t)\} = \Pr\left[\Gamma_{u}(t)<\Gamma_{th}\right]$,
%\begin{equation}
%TOP_{u}\{\vec{q}_{u}(t), j(t)\} = \Pr\left[\Gamma_{u}(t)<\Gamma_{th}\right],
%\end{equation}
where $\Pr$ outputs the probability calculated with respect to (w.r.t.) the aforementioned small-scale fadings. Then, the ergodic outage duration (EOD) of the UAV $u$ travelling with trajectory $\vec{q}_{u}(t), \forall t\in[0, T_{u}]$ from $\vec{q}_{u}(I)$ to $\vec{q}_{u}(D)$ can be expressed as
\begin{equation}
EOD_{u}\{\vec{q}_{u}(t), \hat{i}(t)\}
=\int^{T_{u}}_{0}TOP_{u}\{\vec{q}_{u}(t), \hat{i}(t)\}dt.\label{EOD_{u}}
\end{equation}

According to (\ref{EOD_{u}}), the UAV has more freedom to adjust its flying trajectory for visiting stronger wireless coverage areas (say, regions with lower TOP) if longer flight time budget $T_{u}$ is achievable. However, $T_{u}$ is commonly expected to be as short as possible, for the consideration of energy consumption and time cost for accomplishing the corresponding mission. Therefore, a dilemma of minimizing both $T_{u}$ and $EOD_{u}$ exists inevitably. To balance this, we focus on minimizing the weighted sum of $T_{u}$ and $EOD_{u}\{\vec{q}_{u}(t), \hat{i}(t)\}$ via designing $\vec{q}_{u}(t)$ and $\hat{i}(t)$. Unfortunately, continuous time $t$ implies infinite amount of velocity constraints and location possibilities, leading the UAV path planning task too sophisticated to be handled. Alternatively, the flight period $T_{u}$ is uniformly divided into $N$ time slots, making the navigation task practically trackable. The duration of each time slot $\Delta_{t}=T_{u}/N$ is controlled to be sufficiently small so that the distance, pathloss and antenna gain from each sector towards the UAV can be considered as approximately static within arbitrary time slot.\footnote{In the case of $\Delta_{t}\rightarrow0$, the discrete flight trajectory can accurately approach its continuous counterpart, resulting in extremely heavy computation burden. Therefore, the length of time slot $\Delta_{t}$ should be delicately chosen to achieve satisfactory balance of approximation accuracy and computational complexity.}  Besides, sector assignment is commonly dependent on pathloss to avoid non-stop handover in practice, and thus the associated sector within each time slot is assumed unchanged. Therefore, (\ref{EOD_{u}}) can be approximated as $	EOD_{u}\{\vec{q}_{u}(t), \hat{i}(t)\}
\approx \sum_{n=1}^{N}\Delta_{t}TOP_{u}\{\vec{q}_{u}(n), \hat{i}(n)\}$.
%\begin{equation}
%	EOD_{u}\{\vec{q}_{u}(t), j(t)\}
%	\approx \sum_{n=1}^{N}\Delta_{t}TOP_{u}\{\vec{q}_{u}(n), j(n)\}.\label{EOD_{u}_discrete}
%\end{equation}
With given $\vec{q}_{u}(n)$ and $\hat{i}(n)$ for each time slot, $TOP_{u}\{\vec{q}_{u}(n), \hat{i}(n)\}$ can be obtained via numerical signal measurement at the UAV.\footnote{The closed-form expression of $TOP_{u}\{\vec{q}_{u}(n), \hat{i}(n)\}$ cannot be derived because this paper aims to develop a UAV navigation framework for arbitrary small-scale fading environment and the modelling of $h_{iu}, i\in\{1,2,\cdots,3B\}$ is not specified. Besides, $\Delta_{t}$ (typically on the magnitude of second) is relatively greater than the length of channel coherence block (on the magnitude of millisecond) caused by the small-scale fading. Therefore, $TOP_{u}\{\vec{q}_{u}(n), \hat{i}(n)\}$ can be practically evaluated by numerical measurements on the raw received signals at the UAV.} In this regard, we have 
\begin{equation}
	TOP_{u}\{\vec{q}_{u}(n), \hat{i}(n)\} \simeq \frac{1}{L}\sum_{\iota=1}^{L}ITOP\{\vec{q}_{u}(n), \hat{i}(n)\vert h(\iota)\},\label{approx_TOP}
\end{equation}
where $h(\iota)$ indicates one realization of the involved small-scale fading components, $L$ represents the amount of signal measurements, the TOP indicator $ITOP\{\vec{q}_{u}(n), \hat{i}(n)\vert h(\iota)\} = 1$ if $\Gamma_{u}\{\vec{q}_{u}(n), \hat{i}(n)\vert h(\iota)\}<\Gamma_{th}$ and $ITOP\{\vec{q}_{u}(n), \hat{i}(n)\vert h(\iota)\}=0$ otherwise. Note that $L\gg1$ stands in practice, which means that the approximation (\ref{approx_TOP}) is feasible to be treated as an equation.
Then, the corresponding optimization problem can be stated as 
\begin{subequations}
	%\vspace{-.5cm}
	\begin{align}
	(\text{P}1):\min\limits_{\vec{v}_{u}(n)}&\frac{\tau\Delta_{t}}{L}  \sum_{n=1}^{N}\sum_{\iota=1}^{L}ITOP\{\vec{q}_{u}(n), \hat{i}(n)\vert h(\iota)\} + N,\label{ProposedProblem}\\
	%&\text{s.t.}\nonumber\\
	\text{s.t.}\hspace{.5cm}
	&\hat{i}(n) = \underset{i\in\{1,2,\cdots,3B\}}{\arg\min}PL^{i}\left[\vec{q}_{u}(n)\right], \label{Constraint3}\\
	&\vec{q}(n + 1) = \vec{q}(n) + V_{u}\Delta_{t}\vec{v}_{u}(n), \Vert\vec{v}_{u}(n)\Vert = 1, \label{Constraint4}\\
	&\vec{q}_{\text{lo}}\preceq\vec{q}_{u}(n)\preceq\vec{q}_{\text{up}}, \vec{q}_{u}(0)=\vec{q}_{u}(I), \vec{q}_{u}(N)=\vec{q}_{u}(D), \label{Constraint5}
	%\vspace{-.5cm}
	\end{align}
\end{subequations}
where $\tau$ is the weight balancing the aforementioned minimization dilemma, $V_{u}$ represents the UAV's flying velocity and $\vec{v}_{u}(n)$ specifies the mobility direction. The constraint (\ref{Constraint3}) holds because the sector association strategy is dependent sorely on pathlosses from all the sectors within each time slot and it is clear that the UAV should always pair with the sector which can offer the least degree of pathloss. 

It is straightforward to conclude that antenna gain and LoS/NLoS condition from each sector to the UAV are dependent on the UAV's location with given building and BS distribution, which further impacts the corresponding pathloss and type of small-scale fading. This makes it extremely sophisticated to solve problem (P1) via standard optimization methods, if not impossible.
%, because the considered 3D antenna model, building distribution-based pathloss model, un-specified small-scale fading setup are coupled with each other in a complex manner.  
To provide a better alternative solving the proposed optimization problem (P1), a DRL-aided solution with a novel QiER framework is proposed in this paper.

\section{Deep Reinforcement Learning}
This section is established to give a brief introduction to DRL basics, which is of importance for understanding the proposed DRL-aided solution and the corresponding key notation. For more comprehensive knowledge on RL and DRL, please refer to the celebrated textbook \cite{sutton2018reinforcement}.

The training of RL agent is based on MDP consisting of five components, listed in a tuple $(\mathcal{S},\mathcal{A},\mathcal{T},r,\gamma)$. A state $s_{t}\in\mathcal{S}$ denotes RL agent's observation from the environment at trial $t$. An action $a_{t}\in\mathcal{A}$ represents the agent's choice at trial $t$ following an action selection policy $\mathcal{\pi}(s_{t},a_{t})$. The policy $\pi(s_{t},a_{t}):\mathcal{S}\times\mathcal{A}\rightarrow[0,1]$ claims the probability distribution of picking action $a_{t}$ for state $s_{t}$, constrained by $\sum_{a_{t}\in\mathcal{A}}\pi(s_{t},a_{t})=1$. After executing an action, state transition function $\mathcal{T}\!\!=\!\Pr(s_{t+1}\vert s_{t},a_{t}):\mathcal{S}\times\mathcal{A}\times\mathcal{S}\rightarrow[0,1]$ characterizes state transition $s_{t}\rightarrow s_{t+1}$. 
%If $\mathcal{T}$ is not available, the MDP can still be solved via temporal difference (TD)-based approach, which claims the "model-free" learning progress. 
An immediate reward $r_{t}(s_{t},a_{t})$ acts as performance metric determining how good the selected action $a_{t}$ is, for state $s_{t}$. A scalar factor $\gamma\in[0,1]$ is invoked to discount future rewards, which can help reduce variance caused by the reward function and achieve the convergence of RL algorithms.

While interacting with the environment, the RL agent chooses an action $a_{t}$ for observed state $s_{t}$ at trial $t$ following current action selection policy $\pi(s_{t},a_{t})$. After executing the selected action, state transition $s_{t}\rightarrow s_{t+1}$ occurs and a scalar reward $r_{t}(s_{t},a_{t})$ will be generated. Then, the experience  $exp_{t}=\{s_{t},a_{t},r_{t},s_{t+1}\}$ can be collected to train the RL agent. The state-action value function $Q_{\pi}(s_{t},a_{t})$ (i.e., Q function) derives the accumulated-rewards and reflects the long-term return of acting $a_{t}$ over $s_{t}$ following current action selection policy $\pi$, given by
\begin{equation}
Q_{\pi}(s_{t},a_{t})=\mathbb{E}_{\pi}\left[G_{t}=\sum_{n_{t}=0}^{+\infty}\gamma^{n_{t}}r_{t+n_{t}}\vert s_{t}=s,a_{t}=a\right],
\end{equation}
where $G_{t}$ calculates the discounted accumulated-rewards. The state-action value function $Q_{\pi}(s_{t},a_{t})$ satisfies the Bellman equation, shown as
\begin{equation}
Q_{\pi}(s_{t},a_{t})=\mathbb{E}_{\pi}\left[r_{t}+\gamma\sum_{s_{t+1}\in\mathcal{S}}\mathcal{T}(s_{t+1}\vert s_{t},a_{t})\sum_{a_{t+1}\in\mathcal{A}}\pi(s_{t+1},a_{t+1})Q_{\pi}(s_{t+1},a_{t+1})\right].\label{Q_function_Bellman}
\end{equation}

The RL agent aims to find the optimal policy $\pi^{*}$ which is expected to maximize the long-term return, i.e., $Q^{*}(s_{t},a_{t})=\underset{\pi}{\max}Q_{\pi}(s_{t},a_{t})$. In the case of known optimal Q function $Q^{*}(s_{t},a_{t})$, the optimal action selection policy can be given by $\pi^{*}[s_{t},a_{t}=\underset{a\in\mathcal{A}}{\arg\max}Q^{*}(s_{t},a)]=1$. 

Therefore, an important goal of RL agent is to find the optimal Q function which follows Bellman optimality equation \cite{sutton2018reinforcement}, shown as
\begin{equation}
Q^{*}(s_{t},a_{t})=r_{t}+\gamma\sum_{s_{t+1}\in\mathcal{S}}\mathcal{T}(s_{t+1}\vert s_{t},a_{t})
\underset{a_{t+1}\in\mathcal{A}}{\max}Q^{*}(s_{t+1},a_{t+1}).\label{Optimal_Q_function}
\end{equation}

Unfortunately, (\ref{Optimal_Q_function}) is non-linear and admits no closed-form solution, which can alternatively be solved through iterative algorithms \cite{liu2018deepnap}. Specifically, (\ref{Optimal_Q_function}) can be deduced recursively to achieve the optimality $Q^{*}(s_{t},a_{t})$, via temporal difference (TD) learning. The estimation of Q function can be gradually polished by directly interacting with the environment and sampling the experience sequence $exp_{t}$, which completes the model-free TD learning procedure.  One of popular RL approaches is Q learning which applies the recursive updating rule on Q function $Q(s_{t},a_{t})$, given by
\begin{equation}
Q(s_{t},a_{t})\leftarrow Q(s_{t},a_{t})+
\alpha_{lr}\left[r_{t}+\gamma\underset{a_{t+1}\in\mathcal{A}}{\max}Q(s_{t+1},a_{t+1})-Q(s_{t},a_{t})\right],\label{Recursively_updating_Q_function}
\end{equation}
where $\delta_{t}=r_{t}+\gamma\underset{a_{t+1}\in\mathcal{A}}{\max}Q(s_{t+1},a_{t+1})-Q(s_{t},a_{t})$ represents the TD error and $\alpha_{lr}\in(0,1]$ denotes the learning rate. It is well-known that the optimum $Q^{*}(s_{t},a_{t})$ can be achieved when the state-action pairs are sufficiently experienced and the learning rate is properly chosen \cite{sutton2018reinforcement}.

%\subsection{Tabular RL}
%Tabular RL (e.g., Q learning) is a widely-applied and simple framework to solve MDP-related problems, in which the tabular Q-table is the key component recording the Q values for possible state-action pairs. In tabular RL, the agent is supposed to interact with the environment consistently and converge to the optimality $Q^{*}(s_{t},a_{t})$ via recursively updating $Q(s_{t},a_{t})$ as mentioned in (\ref{Recursively_updating_Q_function}). Unfortunately, this table-based RL method suffers form the curse of dimension, i.e., it becomes inefficient if the state space and/or the action space are huge (or, even continuous). To crack this nut, instead of applying Q-table to store $Q(s_{t},a_{t})$ for each state-action pair, function approximation technique is invoked to approximate the Q function, e.g., linear combinations of features, decision tree, nearest neighbours and artificial neural networks (ANNs). 

\subsection{Deep RL}
To crack the nut rooted from high-dimensional state and/or action spaces, instead of applying Q-table (e.g., tabular RL) to store $Q(s_{t},a_{t})$ for state-action pairs, function approximation technique is used to approximate the Q function, e.g., artificial neural networks (ANNs). Adopting DNN \cite{tao2019learning} (i.e., ANN with deep hidden layers) to approximate Q function is a popular and practical solution \cite{mnih2015human}, i.e.,
%\begin{equation}
$Q(s_{t},a_{t})\approx Q(s_{t},a_{t}\vert\boldsymbol{\theta}), \label{Q_function_approx_theta}$
%\end{equation}
where parameter vector $\boldsymbol{\theta}$ corresponds to the weight coefficients and biases of all links in the DNN. 
%The DNN-aided approximation (\ref{Q_function_approx_theta}) mainly delivers two advantages over tabular RL approach: 1) the ability to predict Q values for inexperienced state-action pairs, which means that it enables generalization; 2) it can drastically reduce the computation complexity because only the parameter vector $\boldsymbol{\theta}$ is necessary to be updated, instead of recording Q values for all experienced state-action pairs. 
The parameter vector $\boldsymbol{\theta}$ can be updated via bootstrapping method to minimize the loss function, defined as
\begin{equation}
\mathcal{L}(\boldsymbol{\theta})=
\left[r_{t}+\gamma\underset{a_{t+1}\in\mathcal{A}}{\max}Q(s_{t+1},a_{t+1}\vert\boldsymbol{\theta}^{-})-Q(s_{t},a_{t}\vert\boldsymbol{\theta})\right]^2, \label{loss_function_w_target}
\end{equation} 
where $Q(s_{t},a_{t}\vert\boldsymbol{\theta}^{-})$ with parameter vector $\boldsymbol{\theta}^{-}$ indicates the target network \cite{mnih2015human},
%The target network $Q(s_{t},a_{t}\vert\boldsymbol{\theta}^{-})$ shares the same structures of the online network $Q(s_{t},a_{t}\vert\boldsymbol{\theta})$, where the updating frequency of $\boldsymbol{\theta}^{-}$ is much less than that of $\boldsymbol{\theta}$. Specifically, the target network will be synchronized to the online network with a given frequency, in terms of updating $\boldsymbol{\theta}^{-}\leftarrow\boldsymbol{\theta}$.  and thus can solve the problem of learning estimates from estimates
which is applied to enhance the stability of learning process \cite{mnih2015human}. 

To further enhance learning performance of deep Q network (DQN), several advanced DRL algorithms were proposed, e.g., double DQN (DDQN) and dueling DQN. Specifically, DDQN approach can help relax the maximization bias brought by $\max$ operation in (\ref{loss_function_w_target}), via allocating action selection and action evaluation into separate networks \cite{van2015deep}. Besides, duelling DQN technique decouples state value and state-dependent action advantages into different streams, which is able to offer better policy evaluation quality, especially for learning tasks containing large amount of similar-valued actions \cite{wang2016dueling}. 

\subsection{Multi-Step Learning}

Standard DRL algorithms apply one-step information to calculate the loss function (\ref{loss_function_w_target}) and train the online network, which may not be adequate and thus lead to poor predictability. Monte Carlo (MC)-based approaches invoke all future state-action pairs to update the online network, but the computation burden could be extremely unbearable. To commit a good balance between one-step learning and MC-aided counterpart, multi-step learning strategy \cite{sutton2018reinforcement} was proposed via taking $N_{ms}$-step-forward knowledge into account. Multi-step learning is prone to help achieve
more satisfactory learning performance, with a delicately chosen step-length $N_{ms}$. Specifically, the $N_{ms}$-step discounted accumulated-reward from a given state $s_{t}$ can be rewritten as
%\begin{equation}
	$r_{t:t+N_{ms}} = \sum_{n_{ms}=0}^{N_{ms} - 1}\gamma^{n_{ms}}r_{t+n_{ms}+1}\label{multistep_return}$.
%\end{equation}
Based on (\ref{loss_function_w_target}), the loss function for $N_{ms}$-step learning can be derived as
\begin{equation}
\mathcal{L}(\boldsymbol{\theta})=
\left[	r_{t:t+N_{ms}} +\gamma^{N_{ms}}\underset{a'\in\mathcal{A}}{\max}Q(s_{t+N_{ms}},a'\vert\boldsymbol{\theta}^{-})-Q(s_{t},a_{t}\vert\boldsymbol{\theta})\right]^2. \label{loss_function_w_target_multistep}\vspace{-.5cm}
\end{equation} 
\subsection{Prioritized Experience Replay}

In the simplest RL framework, the experienced transition $exp_{t}$ is utilized only once and then discarded after the parameter (of policy, value function or model) updating, which brings two shortcomings: 1) inefficient transition sampling, implying that some rare but meaningful transitions might be forgotten rapidly; and 2) highly-correlated transitions, indicating that the independent and identical transition distribution is contaminated. To facilitate the aforementioned disadvantages, experience replay (ER) technique storing experienced transitions into a finite-capacity buffer was proposed \cite{lin1992self, mnih2015human}. Then, a mini-batch of transitions can be sampled to realize the training of DRL agent. The ER makes it possible to break the temporal correlations of experienced transitions via mixing recent and former experiences into the replay buffer, which guarantees that rarely-experienced transitions get fairer chances to be utilized. Through scarifying computation and memory for recording and sampling, ER technique lightens the burden of requiring large amount of experiences for training. However, this compromise is worthy because the interactions between RL agent and environment are more resource-expensive in general \cite{schaul2015prioritized}.

To further improve the efficiency of ER approach, advanced alternative entitled prioritized ER (PER) was proposed \cite{schaul2015prioritized}, in which the recorded experiences are prioritized when being sampled from the ER buffer. The reason why PER method works better is that some transitions are more valuable and meaningful than others for training the agent. While ER technique frees the agent from processing transitions with the order they are experienced, PER can help liberate RL agent from recalling experiences with frequencies proportional to their occurrence probabilities. 
%In this paper, we present a QiER framework for DRL, which can be looked as a PER alternative.
%For realizing prioritization of transitions, priority is associated with each experience, where the priority is usually measured by the corresponding absolute value of TD error. To circumvent the issue of diversity loss in PER, stochastic sampling strategy is adopted, which can achieve a trade-off between greedy prioritization and uniformly-random sampling. Unfortunately, priority-based sampling adopted by PER introduces biases into the updating of agent's DNNs, which can be corrected by weighted importance-sampling (IS) technique reducing the gradient magnitudes for high-priority samples.

\section{Quantum State and Quantum Amplitude Amplification}
In this section, several basic concepts in quantum computation are briefly introduced, which is of essence to the development of DRL-QiER solution.

\subsection{Quantum State}
In quantum mechanics, a quantum state of a closed quantum system can be described by a unit vector in Hilbert space. Specifically, a quantum state $\ket{\Psi_{c}}$ (Dirac notation) comprised of $\hat{n}$ quantum bits (qubits\footnote{A qubit can be realized by a two-state system, e.g., 1) a two-level atom, in which $\ket{0}$ denotes the ground state and $\ket{1}$ indicates the excited state; 2) a photon, where $\ket{0}$ represents the horizontal polarization state and $\ket{1}$ means the vertical polarization state; or 3) a spin system, in which the states of spin up and spin down are described by $\ket{0}$ and $\ket{1}$, respectively.}) can be expressed as \vspace{-.5cm}
\begin{equation}
\ket{\Psi_{c}}= \ket{\Psi_{1}}\otimes\ket{\Psi_{2}}\otimes\dots\otimes\ket{\Psi_{\hat{n}}} = \sum_{p = 00...0}^{\overset{\hat{n}}{\overbrace{11...1}}}h_{p}\ket{p},
\end{equation}
where $\ket{\Psi_{e}},e\in[1,\hat{n}]$ represents the $e$-th qubit, $h_{p}$ means the complex coefficient (i.e., probability amplitude) of eigenstate $\ket{p}$ subject to $\sum_{p = 00...0}^{11...1}\vert h_{p}\vert^{2} = 1$ and $\otimes$ denotes the tensor product. The representation of $\hat{n}$-qubit quantum state $\ket{\Psi_{c}}$ follows the quantum phenomenon known as \textit{state superposition principle}. That is, the $\ket{\Psi_{c}}$ can be regarded as the superposition of $2^{\hat{n}}$ eigenstates, ranged from $\ket{00...0}$ to $\ket{11...1}$. 
As a special case, a two-eigenstate quantum system (say, a single qubit) can be described as an arbitrary superposition state of eigenstates $\ket{0}$ and $\ket{1}$, given by
\begin{equation}
\ket{\Psi} = \alpha\ket{0} + \beta\ket{1},
\label{Psi}
\end{equation}
where the complex coefficients $\alpha=\bra{0}\ket{\Psi}$ and $\beta=\bra{1}\ket{\Psi}$ denote the probability amplitudes for eigenstates $\ket{0}$ and $\ket{1}$, respectively. Note that the single-qubit superposition $\ket{\Psi}$ is a unit vector (i.e., $\braket{\Psi}{\Psi} = 1$) in Hilbert space spanned by orthogonal bases $\ket{0}$ and $\ket{1}$, subject to $\vert\alpha\vert^2 + \vert\beta\vert^2 = 1$. According to \textit{quantum collapse phenomenon}, after measurement or observation of an external experimenter, $\ket{\Psi}$ will collapse from its superposition state onto one of its eigenstates $\ket{0}$ and $\ket{1}$ with probabilities $\vert\alpha\vert^2$ and $\vert\beta\vert^2$, respectively.

\subsection{Quantum Amplitude Amplification}
For a two-eigenstate qubit $\ket{\Psi}$, the probability amplitudes of each eigenstate can be changed via a quantum operation (e.g., Grover iteration\cite{nielsen2010quantum}), gradually modifying the collapse probability distribution. Two unitary reflections are applied to achieve Grover iteration, given by
\begin{equation}
	\boldsymbol{U}_{\ket{0}} = \boldsymbol{I} - (1 - e^{j\phi_{1}})\ket{0}\bra{0},\label{Grover_ket0_initial}
\end{equation}
\begin{equation}
	\boldsymbol{U}_{\ket{\Psi}} = (1 - e^{j\phi_{2}})\ket{\Psi}\bra{\Psi} - \boldsymbol{I}, \label{Grover_ketPsik_initial}
\end{equation}
where $\{\phi_{1},\phi_{2}\}\in[0,2\pi]$, $\boldsymbol{I}$ indicates identity matrix, and $\bra{0}$ and $\bra{\Psi}$ are Hermitian transposes of $\ket{0}$ and $\ket{\Psi}$, respectively. Then, the Grover iterator can be formulated as 
$\boldsymbol{G} = \boldsymbol{U}_{\ket{\Psi}}\boldsymbol{U}_{\ket{0}}$, which remains unitary.
After $m$ times of acting $\boldsymbol{G}$ on $\ket{\Psi}$, the two-eigenstate qubit with updated probability amplitudes can be given by
$\ket{\Psi} \leftarrow \boldsymbol{G}^{m}\ket{\Psi}.$
Two updating approaches can be used to accomplish quantum amplitude amplification task: 1) $m=1$ with dynamic parameters $\phi_{1}$ and $\phi_{2}$; and 2) dynamic $m$ with fixed parameters $\phi_{1}$ and $\phi_{2}$ (e.g., $\pi$). The latter updating method can only change the probability amplitudes in a discrete manner, and thus the former solution is chosen in this paper.
%, i.e., Grover iteration with flexible parameters $\phi_{1}$ and $\phi_{2}$. However, the later updating method will be simulated in the numerical results section to serve as a baseline.
\begin{prop}
	\label{prop_G_Psi}
	For Grover iteration with flexible parameters, the overall effects of $\boldsymbol{G}$ on the superposition $\ket{\Psi}$ can be derived analytically as
	$\boldsymbol{G}\ket{\Psi}=(\mathcal{Q}-e^{j\phi_{1}})\alpha\ket{0} + (\mathcal{Q} - 1) \beta\ket{1}$,
	where $\mathcal{Q} = (1 - e^{j\phi_{2}})\left[1 - (1 - e^{j\phi_{1}})\vert \alpha\vert^2\right]$ and $\vert(\mathcal{Q}-e^{j\phi_{1}})\vert^2\vert\alpha\vert^2 + \vert(\mathcal{Q} - 1)\vert^2\vert\beta\vert^2 = 1$.
\end{prop}
\begin{IEEEproof}
	The effects of $\boldsymbol{U}_{\ket{0}}$ on $\ket{0}$ and $\ket{1}$ are expressed as
	\begin{equation}
		\boldsymbol{U}_{\ket{0}}\ket{0}=\left[ \boldsymbol{I} - (1 - e^{j\phi_{1}})\ket{0}\bra{0}\right]\ket{0}=e^{j\phi_{1}}\ket{0},
	\end{equation}
	\begin{equation}
		\boldsymbol{U}_{\ket{0}}\ket{1}=\left[ \boldsymbol{I} - (1 - e^{j\phi_{1}})\ket{0}\bra{0}\right]\ket{1}=\ket{1},
	\end{equation}
	respectively.
	Then, we obtain 
	\begin{equation}
		\boldsymbol{U}_{\ket{0}}\ket{\Psi}=\left[ \boldsymbol{I} - (1 - e^{j\phi_{1}})\ket{0}\bra{0}\right]\ket{\Psi} =e^{j\phi_{1}}\alpha\ket{0} + \beta\ket{1},
	\end{equation}
	where $\boldsymbol{U}_{\ket{0}}$ plays the role as a \textit{conditional phase shift operator}. 
	
	Furthermore, we get
	\begin{align}
		\boldsymbol{G}\ket{\Psi}=\boldsymbol{U}_{\ket{\Psi}}\boldsymbol{U}_{\ket{0}}\ket{\Psi}&=(1 - e^{j\phi_{2}})\left[\alpha\ket{0}+\beta\ket{1}\right]\left[\alpha^{\dagger}\bra{0}+\beta^{\dagger}\bra{1}\right]\boldsymbol{U}_{\ket{0}}\ket{\Psi} - \boldsymbol{U}_{\ket{0}}\ket{\Psi}\nonumber\\
		&=(\mathcal{Q}-e^{j\phi_{1}})\alpha\ket{0} + (\mathcal{Q} - 1) \beta\ket{1},
		\label{GPsi}
	\end{align}
	where $\mathcal{Q} = (1 - e^{j\phi_{2}})(e^{j\phi_{1}}\vert \alpha\vert^2 + \vert \beta\vert^2) = (1 - e^{j\phi_{2}})\left[1 - (1 - e^{j\phi_{1}})\vert \alpha\vert^2\right]$.
	
	Because Grover operator $\boldsymbol{G}$ is unitary, the updated superposition $\ket{\Psi}\leftarrow\boldsymbol{G}\ket{\Psi}$ still follows the normalization rule of probability amplitudes, i.e., $\vert(\mathcal{Q}-e^{j\phi_{1}})\vert^2\vert\alpha\vert^2 + \vert(\mathcal{Q} - 1)\vert^2\vert\beta\vert^2 = 1$.
	\begin{comment}
			\begin{align}
		\mathcal{Q} &= (1 - e^{j\phi_{2}})(e^{j\phi_{1}}\vert \alpha_{k}\vert^2 + \vert \beta_{k}\vert^2) \nonumber \\
		%&=(1 - e^{j\phi_{2}})(e^{j\phi_{1}}\vert h_{i}\vert^2 + 1 - \vert h_{i}\vert^2) \nonumber \\
		& = (1 - e^{j\phi_{2}})\left[1 - (1 - e^{j\phi_{1}})\vert \alpha_{k}\vert^2\right].
		\end{align} 
	\end{comment}
	%This concludes the proof of {\it Theorem \ref{thmGonAl}}.
\end{IEEEproof}
\begin{cor}\label{cor_symmetric_Grover_general_Psi}
	The ratio between collapse probabilities of $\ket{\Psi} \rightarrow \ket{0}$ before and after being impacted by $G$ can be given by 
	$\vert\mathcal{R}\vert^2\!=\!\vert(1 \!-\! e^{j\phi_{1}} \!-\! e^{j\phi_{2}}) \!-\! (1 \!-\! e^{j\phi_{1}})(1 \!-\! e^{j\phi_{2}})\vert \alpha\vert^2\vert^2, \label{Updated_Probability_Psi}$
	which is symmetric w.r.t. $\phi_{1}=\phi_{2}$ and $\phi_{1}=2\pi-\phi_{2}$.
	Then, the updated collapse probabilities onto eigenstates $\ket{0}$ and $\ket{1}$ can be given by $\vert\mathcal{R}\vert^2\vert \alpha\vert^2$ and $1 - \vert\mathcal{R}\vert^2\vert \alpha\vert^2$, respectively.
\end{cor}
\begin{IEEEproof}
	Based on (\ref{Psi}) and (\ref{GPsi}), the ratio between the probability amplitudes of $\ket{0}$ after being acted by $\boldsymbol{G}$ and before that can be derived as
	$\mathcal{R} 
	=(1 - e^{j\phi_{1}} - e^{j\phi_{2}}) - (1 - e^{j\phi_{1}})(1 - e^{j\phi_{2}})\vert \alpha\vert^2$, which completes the proof.
%	Then, (\ref{Updated_Probability_Psi}) can be derived straightforwardly.
\end{IEEEproof}
\begin{remrk}
	\label{remrk_visualization_Bloch}
	The process of $\ket{\Psi}\leftarrow\boldsymbol{G}\ket{\Psi}$ can be depicted geometrically on the Bloch sphere. In Fig. \ref{Grover rotation on Psi}, $\ket{\Psi}$ is reconstructed in Polar coordinates, given by
	\begin{equation}
		\ket{\Psi}=e^{j\zeta}(\cos\frac{\theta}{2}\ket{0} + e^{j\varphi}\sin{\frac{\theta}{2}}\ket{1})\simeq \cos\frac{\theta}{2}\ket{0} + e^{j\varphi}\sin{\frac{\theta}{2}}\ket{1},\label{Psi_k_Polar_coordination}
	\end{equation}
	%&= h_{i}\ket{a_{i}}+h_{a_{i}^{\perp}}\ket{a_{i}^{\perp}} \nonumber \\
	where $e^{j\zeta}$ poses no observable effects \cite{li2020quantum}. Then, the unit vector $\ket{\Psi}$ on the Bloch sphere is uniquely specified by angle variables $\theta\in[0,\pi]$ and $\varphi\in[0,2\pi)$. 
	The effect of $\boldsymbol{U}_{\ket{0}}$ can be regarded as a clockwise rotation around the $z$-axis by $\phi_{1}$ (the red circle) on the Bloch sphere, leading to the rotation from $\ket{\Psi}$ to $\ket{\Psi'}$. In a similar manner, when the basis is changed from $\{\ket{0},\ket{1}\}$ to $\{\ket{\Psi}, \ket{\Psi^{\perp}}\}$, $\boldsymbol{U}_{\ket{\Psi}}$ results in a clockwise rotation around the new $z$-axis $\ket{\Psi}$ by $\phi_{2}$ (the blue circle), rotating $\ket{\Psi'}$ to $\ket{\Psi^{(1)}}$. Hence, the overall impact of $\boldsymbol{G}$ on $\ket{\Psi}$ is a two-step process rotating the polar angle $\theta$, on the perspective of basis $\{\ket{0},\ket{1}\}$. With flexible $\phi_{1}$ and $\phi_{2}$, it is possible to achieve arbitrary parametric rotation on the Bloch sphere, which serves as the foundation for quantum amplitude amplification task. The smaller $\theta$ is, the higher probability $\ket{\Psi}$ will collapse onto $\ket{0}$ when it is observed by an external examiner, and vice versa. 
\begin{comment}
		\begin{figure}
	\centering
	\vspace{-1cm}
	\includegraphics[width=0.7 \textwidth]{QuantumExp_Psi.png}
	%		\captionsetup{font={scriptsize}}
	\caption{Geometric explanation of the Grover rotation}
	\label{Fig_GeometricExplation_Psi_Initial}
	\vspace{-.7cm}
	\end{figure}
\end{comment}
\begin{figure}[htbp]
	\centering  
	\subfigcapskip = -.3cm
	\vspace{-.7cm}
	\subfigure[\tiny Grover rotation on $\ket{\Psi}$]{
		\label{Grover rotation on Psi}
		\includegraphics[scale=0.15]{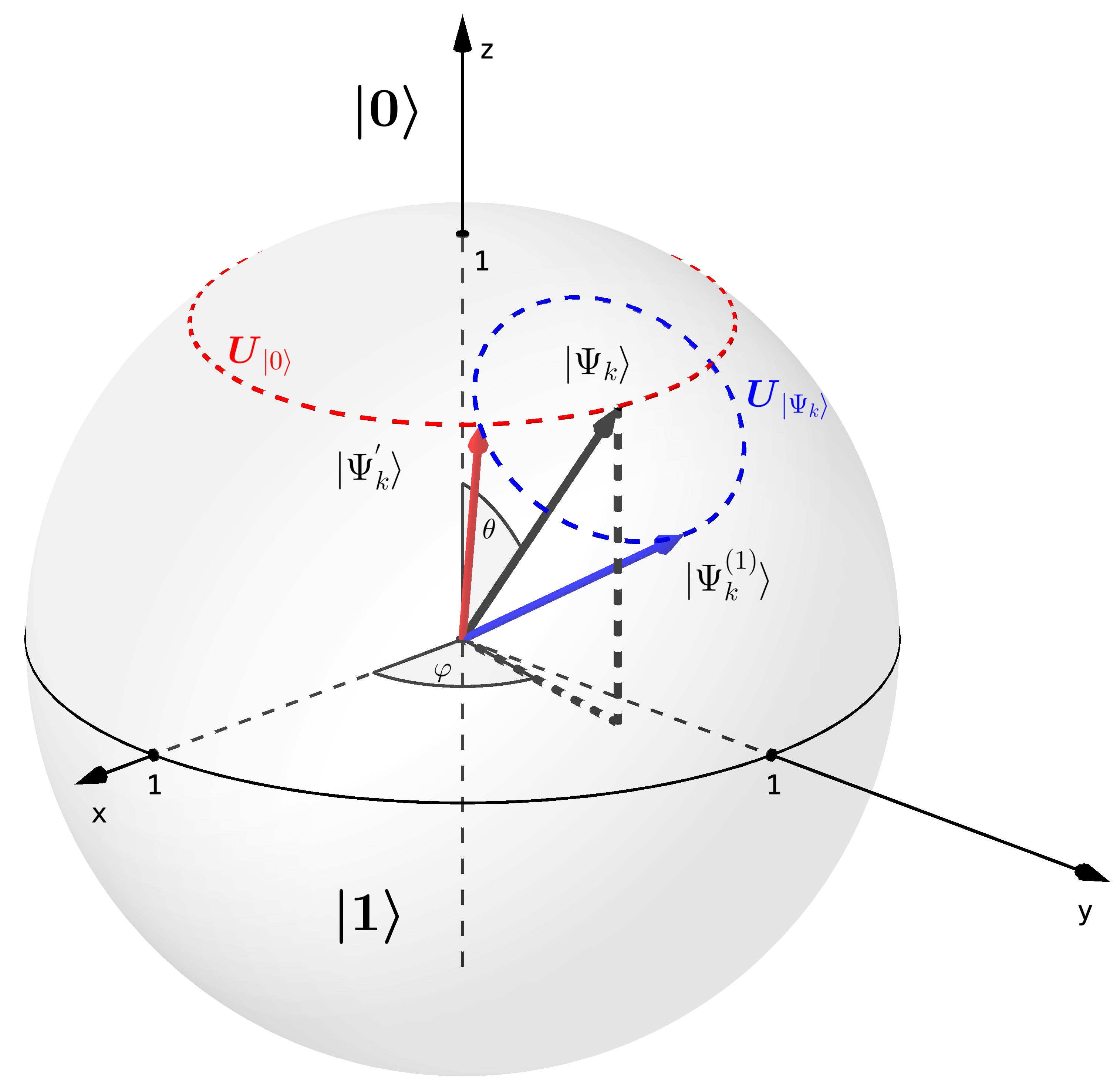}}\hspace{3cm}
	\subfigure[\tiny Grover rotation on $\ket{+}$]{
		\label{Grover rotation on +}
		\includegraphics[scale=0.15]{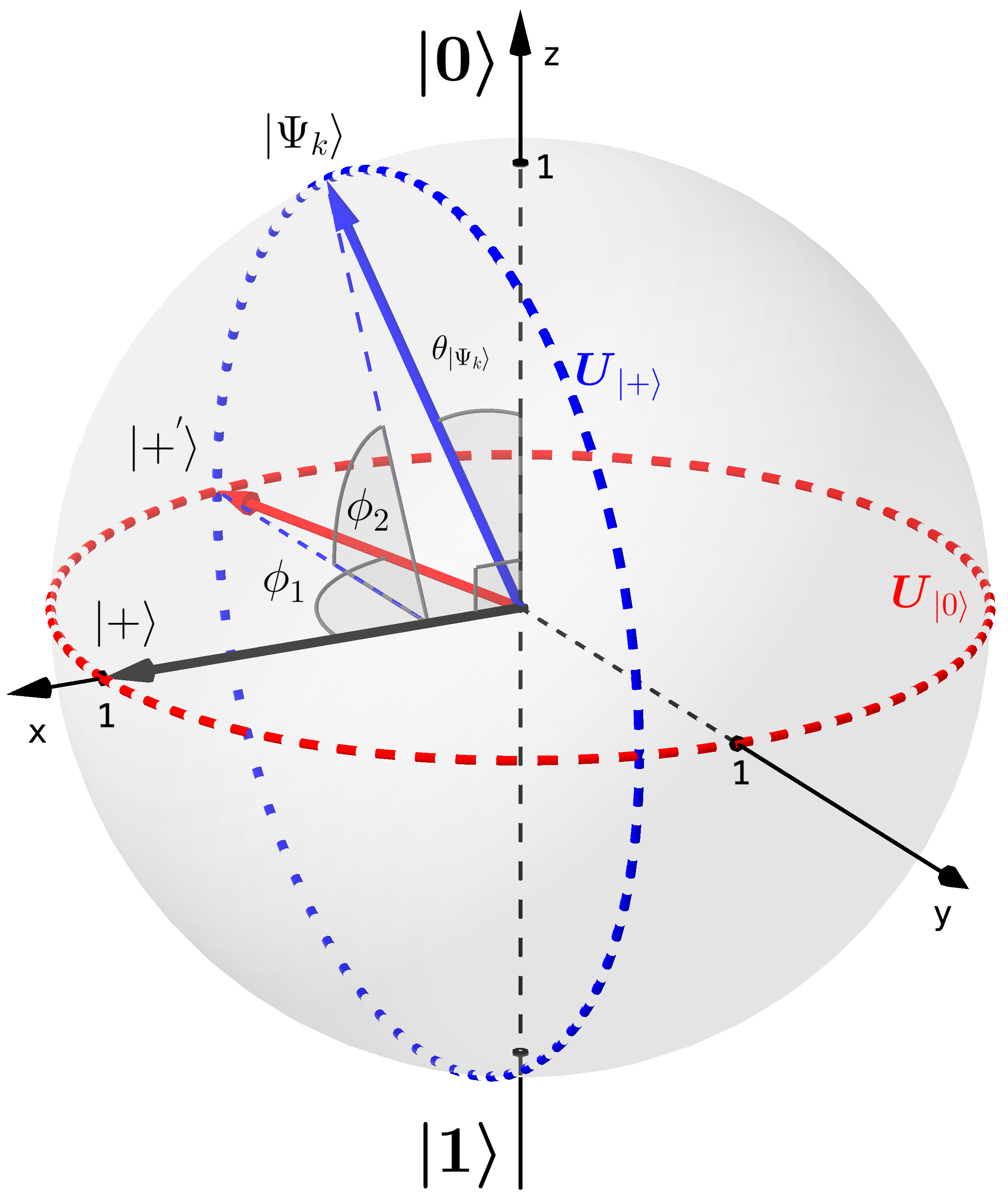}}
	\captionsetup{font={scriptsize}}
	\vspace{-.5cm}
	\caption{Geometric explanation of the Grover rotation}
	\vspace{-1cm}
	\label{Geometric explanation of the Grover rotation}
\end{figure}
\end{remrk}

\section{DRL-QiER Algorithm}
In this section, a DRL-QiER solution is developed to solve optimization problem (P1).
\vspace{-.5cm}
\subsection{The MDP Formulation}
\label{MDP_DRL_QiER}
To solve the optimal trajectory planning problem (P1) via DRL-aided technique, the first step is to map it into an MDP, which can be described as follows.
\begin{itemize}
	\item $\mathcal{S}$: The state space consists of possible UAV locations $\vec{q}_{u}$ under constraint $\vec{q}_{\text{lo}}\preceq\vec{q}_{u}\preceq\vec{q}_{\text{up}}$, which means that the state space is continuous.
%	the state refers to the UAV's coordinate $\vec{q}_{u}(n)\in\mathcal{S},\forall n\in[1,N]$, where 
	\item $\mathcal{A}$: The continuous action space involves all the feasible flying directions $\vec{v}_{u}$ under constraint $\Vert\vec{v}_{u}\Vert=1$. To break the curse of dimensionality caused by continuous state and action spaces, the action space is discretized as $\mathcal{A}=\left\{[1,0,0], [0,1,0], [-1,0,0], [0,-1,0], [\sqrt{2}/2,\sqrt{2}/2,0],\right.\\\left. [-\sqrt{2}/2,\sqrt{2}/2,0], [\sqrt{2}/2,-\sqrt{2}/2,0], [-\sqrt{2}/2,-\sqrt{2}/2,0]\right\}$,
	 corresponding to flying directions right, forward, left, backward, right-forward, left-forward, right-backward and left-backward, respectively. Therefore, the action space contains $N_{fd}=8$ direction options. 
	\item $\mathcal{T}$: The state transition is deterministic and controlled by the mobility constraint (\ref{Constraint4}).
	\item $r$: Our goal is to minimize the weighted sum of time cost and EOD. Thus, we may design the reward function as 
	$r(\vec{q}_{u}) = -1 - \frac{\tau\Delta_{t}}{L}\sum_{\iota=1}^{L}ITOP\{\vec{q}_{u}\vert h(\iota)\}$. The formulation of $r(\vec{q}_{u})$ can be interpreted as follows: 1) for each time of state transition, the agent will receive a movement penality 1, encouraging the UAV to use less steps to generate the trajectory; and 2) on top of the movement penality, the UAV will get a weighted outage duration penality $\frac{\tau\Delta_{t}}{L}\sum_{\iota=1}^{L}ITOP\{\vec{q}_{u}\vert h(\iota)\}$ as well, pushing the UAV to visit locations with stronger wireless coverage quality. Besides, two special cases are considered as follows: 1) once the UAV reaches the predefined destination $\vec{q}_{u}(D)$, the training episode terminates and a positive value $r_{D}$ will replace the reward function; and 2) once the UAV crashes onto the boundary of the considered airspace, the training episode terminates and a negative value $r_{ob}$ will replace the reward function instead. In summary, the aforementioned design of reward function aims to encourage the UAV to reach $\vec{q}_{u}(D)$ with as fewer steps as possible, while avoiding hitting the boundary and visiting areas with weak wireless coverage strength.
	\item $\gamma$: To connect the objective function of (P1) and the discounted accumulated-rewards over each learning episode, the discount factor is chosen as $\gamma=1$.
\end{itemize}

%Note that the formulated MDP is episodic, which means that each training episode terminates once the terminal state is reached and then a new episode will be initiated with the environment being reset. The terminal state of our MDP corresponds to the predefined destination or collision with the boundary.

\subsection{Quantum-Inspired Representation of Experience's Priority}
In the proposed DRL-QiER solution, the priority of experienced transition $exp_{t}$ is represented by the $k$-th qubit, where the scalar index $k$ indicates this transition's location index in the QiER buffer. Specifically, the quantum representation of stored transition's priority can be given by 
\begin{equation}
\ket{\Psi_{k}} = \alpha_{k}\ket{0} + \beta_{k}\ket{1}, \label{quantum_experience_representation}
\end{equation}
where the complex-valued probability amplitudes $\alpha_{k}$ and $\beta_{k}$ follow the normalization constraint $\vert\alpha_{k}\vert^2 + \vert\beta_{k}\vert^2 = 1$. It is worth noting that the eigenstates $\ket{0}$ and $\ket{1}$ in (\ref{quantum_experience_representation}) mean accepting and denying this transition, respectively. After quantum measurement, the superposition $\ket{\Psi_{k}}$ will collapse onto eigenstate $\ket{0}$ with probability $\vert\braket{0}{\Psi_{k}}\vert^2=\vert\alpha_{k}\vert^2$ or eigenstate $\ket{1}$ with probability $\vert\braket{1}{\Psi_{k}}\vert^2=\vert\beta_{k}\vert^2$. The complex coefficients $\alpha_{k}$ and $\beta_{k}$ are of importance and essence in the QiER system, influencing the occurrence probability of accepting or denying the corresponding transition when $\ket{\Psi_{k}}$ is observed. The quantum representation $\ket{\Psi_{k}}$ establishes a bridge between quantum eigenstates and accepting or denying particular transition, which allows us to apply quantum amplitude amplification to realize manipulation of quantum collapse. 

\subsection{QiER Framework}
The proposed QiER framework consists of the following three phases.
\subsubsection{Quantum Initialization Phase} When transition $exp_{t}$ is stored into the QiER buffer with finite capacity $C$, a label $k\in\{1,\dots,C\}$ will be assigned to $exp_{t}$, which specifies the location of $exp_{t}$ being recorded within the QiER buffer.\footnote{The QiER buffer is designed to be with fixed-size capacity in line with standard ER technique of DRL, which means that the first stored experience will be popped out first to create space for recording the new-coming transition when the QiER buffer is fully exploited. Therefore, each recorded experience is supposed to remain in the buffer for a fixed time.} Then, experience $exp_{t}$ and the $k$-th qubit $\ket{\Psi_{k}}$ together will be stored into the QiER buffer, which can be regarded as a collection of $\left(exp_{t},\ket{\Psi_{k}}\right)$. When a new transition is recorded into the QiER buffer and before being sampled out to feed the training agent, its associated qubit $\ket{\Psi_{k}}$ should be initialized as eigenstate $\ket{0}$, i.e., $\ket{\Psi_{k}} \leftarrow \ket{0}$. The reason is that the agent has never been trained with these un-sampled transitions that may have unimaginable potentials to help the agent learn the characteristics of environment with which the agent is interacting. Thus, we set these newly-recorded transitions with the highest priority, encouraging the agent to more likely learn from them.

\subsubsection{Quantum Preparation Phase} After an experience is sampled from the QiER buffer to train the agent, the quantum preparation phase should be performed on its associated qubit, updating the corresponding priority. This is due to two reasons: 1) the TD error of this transition is updated; and 2) the experience becomes older for the agent. 

The uniform quantum state is defined as 
\begin{equation}
	\ket{+} = \frac{\sqrt{2}}{2}\left(\ket{0} + \ket{1}\right), \label{quantum_experience_initial_quantum_state}
\end{equation}
which can be understood as a unit vector on the x-axis of Bloch sphere (Fig. \ref{Grover rotation on +}) with $\theta = \pi/2$ and $\varphi = 0$. The absolute value of TD error $\vert\delta_{t}\vert$ is chosen to reflect priority of the corresponding transition $exp_{t}$. Once a recorded transition is sampled, its associated qubit $\ket{\Psi_{k}}$ should first be reset to the uniform quantum state, i.e., $\ket{\Psi_{k}} \leftarrow \ket{+}$. Then, to map the updated priority of $exp_{t}$ into $\ket{\Psi_{k}}$, one time of Grover iteration with flexible parameters will be applied on the uniform quantum state, shown as
\begin{equation}
	\ket{\Psi_{k}} = \boldsymbol{U}_{\ket{+}}\boldsymbol{U}_{\ket{0}}\ket{+}
	\overset{(a)}{=} (\mathcal{P}-e^{j\phi_{1}})\frac{\sqrt{2}}{2}\ket{0} + (\mathcal{P} - 1) \frac{\sqrt{2}}{2}\ket{1},
\end{equation}
where $\mathcal{P}  = (1 - e^{j\phi_{2}})\left[1 - 0.5(1 - e^{j\phi_{1}})\right]$ and the derivation $(a)$ is based on \textbf{\textit{Proposition}} \textit{\ref{prop_G_Psi}}. According to \textbf{\textit{Remark}} \textit{\ref{remrk_visualization_Bloch}}, the transformation from $\ket{+}$ to $\ket{\Psi_{k}}$ can be depicted on the Bloch sphere as Fig. \ref{Grover rotation on +}. In this example, the phase shift parameters are set as $\phi_{1}<\pi/2$ and $\phi_{2}<\pi/2$. It is straightforward to observe that the probability of collapsing onto eigenstate $\ket{0}$ enlarges after the quantum preparation phase (i.e., $\ket{+}\overset{\boldsymbol{U}_{\ket{+}}\boldsymbol{U}_{\ket{0}}}{\longrightarrow}\ket{\Psi_{k}}$), because the polar angle rotates from $\angle90^{\circ}$ (of $\ket{+}$) to an acute angle $\theta_{\Psi_{k}}$ (of $\ket{\Psi_{k}}$). Similarly, the collapse probability onto eigenstate $\ket{0}$ after one time of Grover iteration on $\ket{+}$ can be kept unchanged or shrinked via selecting feasible combination of phase shift parameters $\phi_{1}\in[0,2\pi]$ and $\phi_{2}\in[0,2\pi]$.

In practical applications, some experiences may be sampled for training with undesired high frequency, leading to over-training issue. Besides, the finite size of QiER buffer could further deteriorate this disservice \cite{de2015importance}, which will cause unfair and biased sampling performance. To circumvent this issue, the replay time of each stored transition should be taken into consideration for the quantum preparation phase, which enables it to enrich sample diversity to improve the learning performance. In the early stage of training the agent, the importance of each experience is ambiguous. However, alongside the learning process, the absolute TD errors of some transitions remain relatively large, despite many times they have been sampled for training. Hence, it is necessary to relate training episode to the quantum preparation phase.

The quantum preparation phase aims to modify the collapse probability onto eigenstate $\ket{0}$, via one time of Grover iteration with free parameters $\phi_{1}$ and $\phi_{2}$. To quantify the amplification step of quantum preparation phase, we let 
\begin{equation}
\phi_{1} = \frac{e^{\frac{\vert\delta_{t}\vert\pi}{\delta_{\text{max}}}} - e^{-\frac{\vert\delta_{t}\vert\pi}{\delta_{\text{max}}}}}{e^{\frac{\vert\delta_{t}\vert\pi}{\delta_{\text{max}}}} + e^{-\frac{\vert\delta_{t}\vert\pi}{\delta_{\text{max}}}}}\frac{\pi}{2} = \frac{\pi}{2}tanh\left(\frac{\vert\delta_{t}\vert\pi}{\delta_{\text{max}}}\right)\in\left[0,\frac{\pi}{2}\right),\label{quantum_amplification_step_phi_1}
\end{equation}
\begin{equation}
\phi_{2} = \frac{rt_{k}}{rt_{\text{max}}}\frac{te}{te_{\text{max}}}\pi + \frac{\pi}{2}\in\left(\frac{\pi}{2}, \frac{3\pi}{2}\right].\label{quantum_amplification_step_phi_2}
\end{equation}  
With (\ref{quantum_amplification_step_phi_1}) and (\ref{quantum_amplification_step_phi_2}), the quantum amplitude amplification is related with the corresponding absolute TD error $\vert\delta_{t}\vert$, maximum TD error $\delta_{\text{max}}$, replay times $rt_{k}$, maximum replay time $rt_{\text{max}}$, current training episode $te$ and the total training episode $te_{\text{max}}$, which means that the quantum preparation phase updates the priority of $exp_{t}$ into its associated $k$-th qubit $\ket{\Psi_{k}}$. 

\begin{remrk}
	The collapse probability of $\ket{\Psi_{k}}$ onto eigenstate $\ket{0}$ versus $\phi_{1}\in[0,2\pi]$ and $\phi_{2}\in[0,2\pi]$ is depicted in Fig. \ref{Collapsing Probability onto ket 0}. From this figure, we can find that $\vert\braket{0}{\Psi_{k}}\vert^2 = 0.5\vert\mathcal{P}-e^{j\phi_{1}}\vert^2$ is a symmetric function w.r.t. $\phi_{1} = \phi_{2}$ and $\phi_{1} = 2\pi - \phi_{2}$, which is a specific case (i.e., $\vert\alpha\vert^2=0.5$) of \textbf{\textit{Corollary}} \textit{\ref{cor_symmetric_Grover_general_Psi}}. If we concentrate on surface within $\phi_{1}\in[0,\pi/2]$ and $\phi_{2}\in[\pi/2, 3\pi/2]$, it is straightforward to conclude that (\ref{quantum_amplification_step_phi_1}) and (\ref{quantum_amplification_step_phi_2}) together can control the quantum amplification step and direction. Specifically, larger $\phi_{1}$ will lead to greater amplitude amplification step, for arbitrary fixed $\phi_{2}$. Besides, $\phi_{2}$ controls the amplification direction, where $\phi_{2}\in[\pi/2,\pi)$ means that the probability of collapsing onto $\ket{0}$ will be enlarged, while $\phi_{2}\in(\pi, 3\pi/2)$ indicates that the probability of collapsing onto $\ket{0}$ will be reduced.
	\begin{figure}[htbp]
		\centering  
		\subfigcapskip = -.5cm
		\vspace{-.7cm}
		\subfigure[\tiny 3D View]{
			\label{3D_View}
			\includegraphics[scale=0.3]{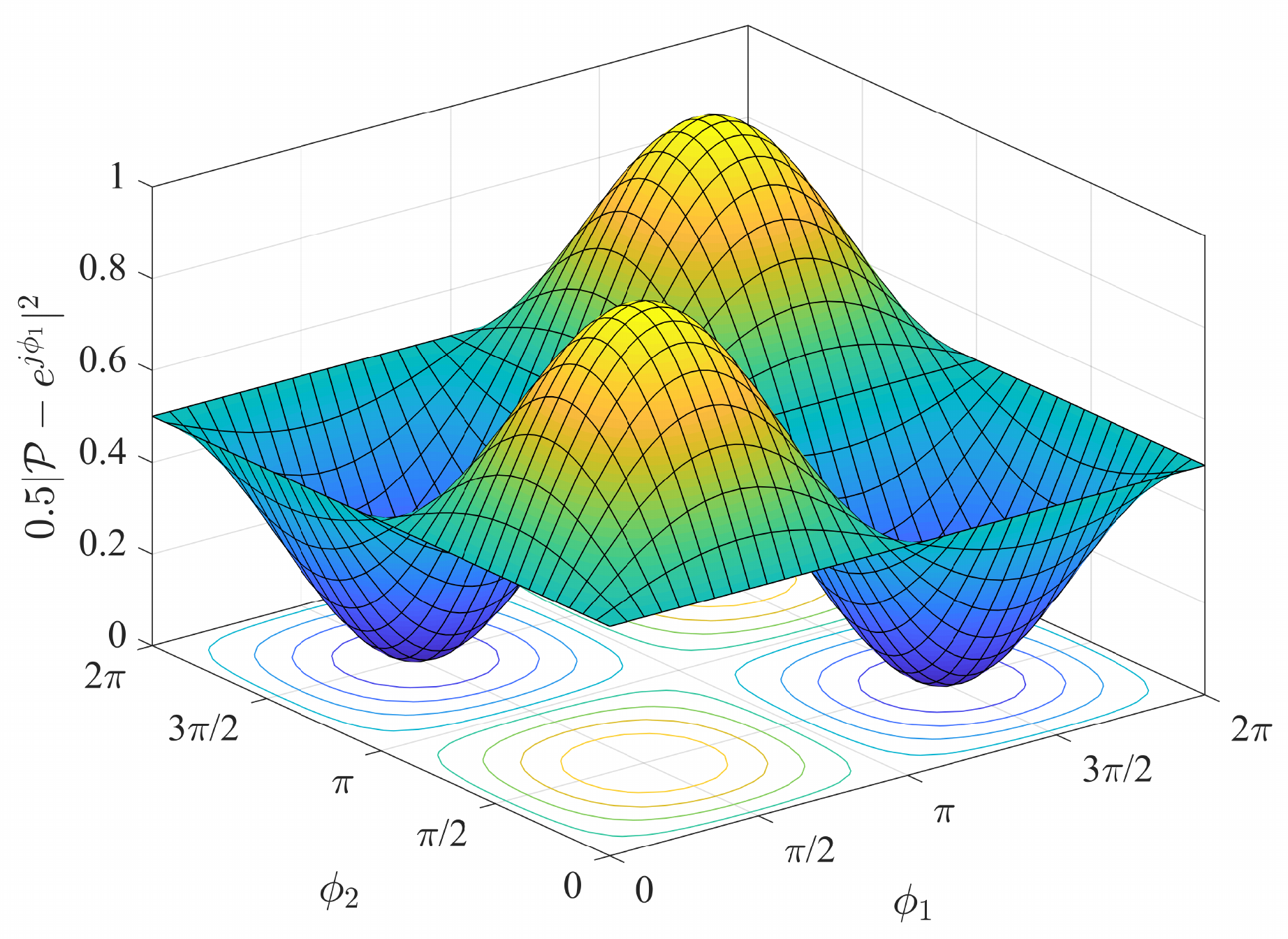}}\hspace{3cm}
		\subfigure[\tiny The corresponding top view]{
			\label{2D_View}
			\includegraphics[scale=0.25]{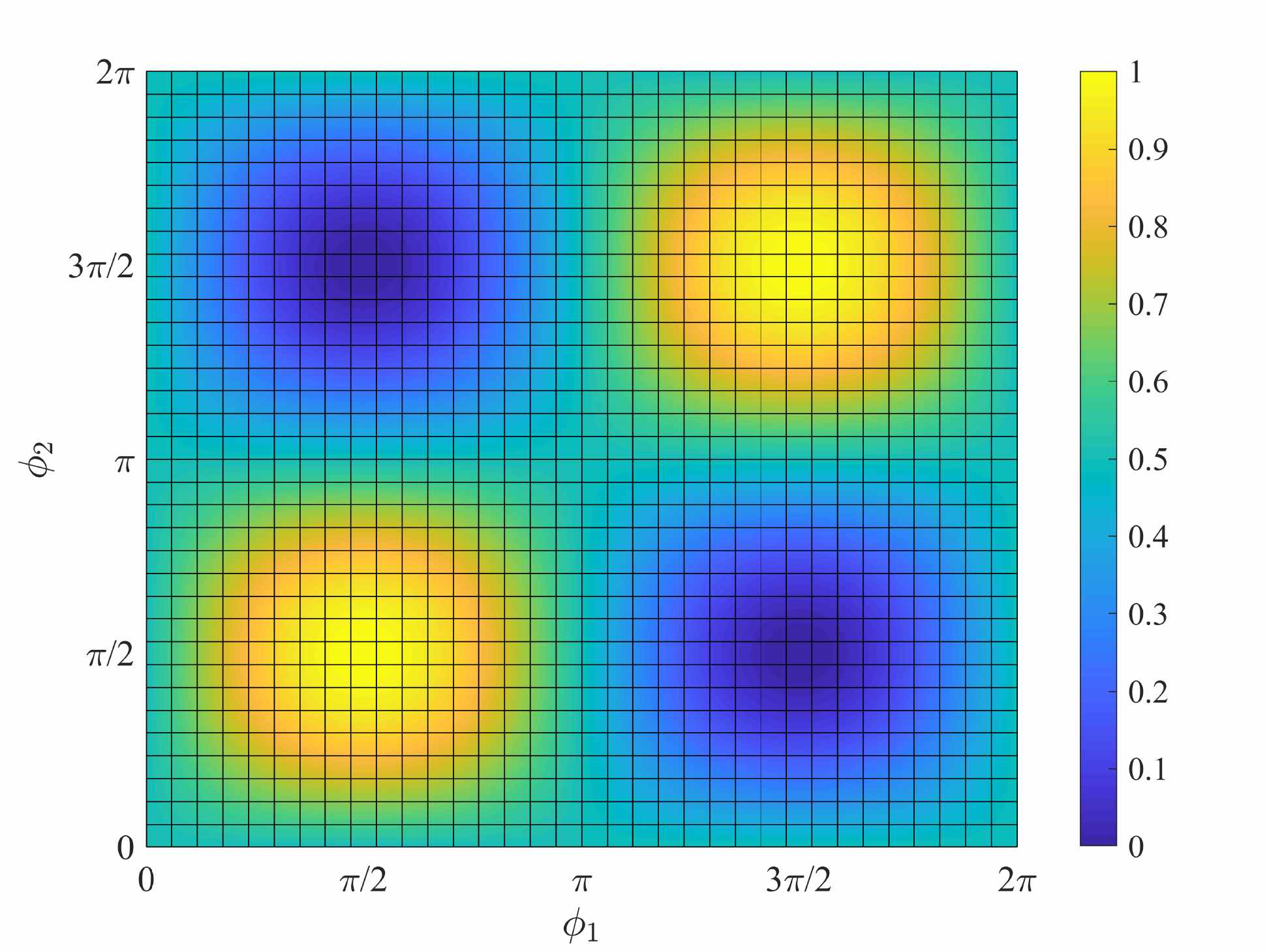}}
		\captionsetup{font={scriptsize}}
		\vspace{-.5cm}
		\caption{Collapse Probability onto $\ket{0}$ versus $\phi_{1}$ and $\phi_{2}$}
		\vspace{-.5cm}
		\label{Collapsing Probability onto ket 0}
	\end{figure}
\end{remrk}

\begin{remrk}
	In the early stage of training, the radio $rt_{k}/rt_{\text{max}}$ remains relatively large because $rt_{\text{max}}$ is not sufficiently updated yet. To avoid unreasonably denying all the sampled transitions in the early stage of training, we introduce the factor $te/te_{\text{max}}$ to steer parameter $\phi_{2}$ in (\ref{quantum_amplification_step_phi_2}).
\end{remrk}

%After performing the initialization and preparation phases, the priority of each stored experience can be determined via quantum measurement on its corresponding qubit, which is the foundation for mini-batch sampling in the proposed DRL-QiER solution. 
\begin{comment}
	\begin{figure}
	\centering
	\includegraphics[width=0.7 \textwidth]{QuantumExp_Psik_from_+.png}
	%		\captionsetup{font={scriptsize}}
	\caption{Geometric explanation of the Grover rotation}
	\label{Fig_GeometricExplation_from_+_to_Psi_k}
	\end{figure}
\end{comment}
%\subsubsection{Depreciation Phase}
%After the preparation phase, the probability of selecting each recorded experience is amplified, which is proportional to the transition's corresponding priority. 
\subsubsection{Quantum Measurement Phase}
After the QiER buffer is fully occupied by recorded transitions, a mini-batch of experiences will be sampled to perform network training for the agent, via standard gradient descent method. To prepare the mini-batch sampling procedure under constraint of priorities, quantum measurement on the associated qubits should be accomplished first. Specifically, the probability of the $k$-th qubit collapsing onto eigenstate $\ket{0}$ can be calculated as $\vert\braket{0}{\Psi_{k}}\vert^2$. Then, the probability of the corresponding experience being picked up during the mini-batch sampling process can be defined as
$bp_{k} = \vert\braket{0}{\Psi_{k}}\vert^2/\sum_{e=1}^{C}\vert\braket{0}{\Psi_{e}}\vert^2$,
%\begin{equation}
%	bp_{k} = \frac{\vert\braket{0}{\Psi_{k}}\vert^2}{\sum_{i=1}^{C}\vert\braket{0}{\Psi_{i}}\vert^2},\label{being_picked_pro_mini_batch}
%\end{equation}
in which the denominator means the sum of collapse probabilities onto eigenstate $\ket{0}$ of qubits that are associated with all stored experiences.

During the mini-batch sampling period, several times of picking recorded experiences from the QiER buffer will be executed, following the generated picking probability vector $\vec{bp}=[bp_{1},bp_{2},
\dots,bp_{C}]$ after quantum measurement phase. Note that the total sampling time is equal to the size of mini-batch, which will be specified in the numerical result section later. 

\begin{remrk}
	Although the QiER buffer involves quantum representations and operations, the corresponding processes, i.e., the quantum initialization phase, the quantum preparation phase and the quantum measurement phase, can be imitated on conventional computing devices without implementing real quantum computations on practical quantum computers. 
\end{remrk}
\begin{remrk}
	The associated qubit of sampled experience should be reset to the uniform quantum state, which means that the corresponding quantum preparation phase starts from the uniform quantum state rather than the previous counterpart. This is in line with the quantum phenomenon where a quantum system will collapse onto one of its eigenstates after an observation. Note that the sampled transitions are still remained in the QiER buffer until they are discarded.
\end{remrk}
\subsection{The Proposed DRL-QiER Solution}
\begin{figure}[htbp]
	\centering\vspace{-1cm}
	\includegraphics[scale=0.63]{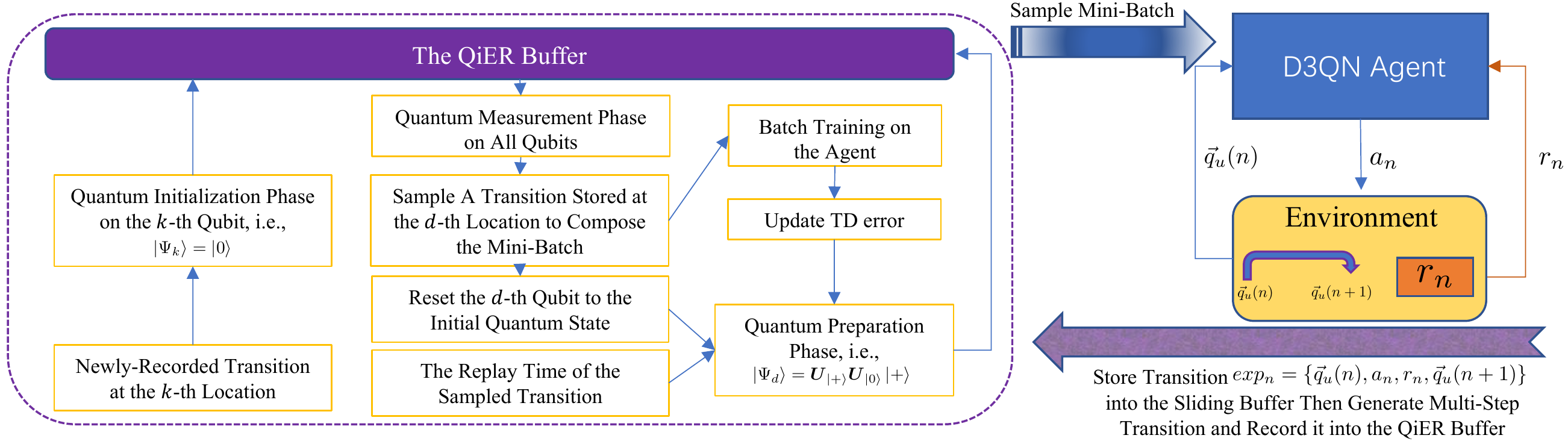}	\captionsetup{font={scriptsize}}
	\vspace{-0.3cm}
	\caption{Flow chart of the proposed DRL-QiER algorithm}
	\label{QuantumExp_FlowChart}
	\vspace{-1cm}
\end{figure}

The proposed DRL-QiER algorithm is summarized in \textbf{Algorithm} \textbf{1}, and its flow chart is illustrated in Fig. \ref{QuantumExp_FlowChart}. To solve the formulated MDP in Section \ref{MDP_DRL_QiER}, DDQN with duelling architecture (D3QN) is adopted to approximate the Q function $Q(\vec{q}_{u}, \vec{v}_{u})$. 
%Compared to the original DQN without duelling layer, duelling DQN offers more efficient and robust learning performance on approximating the Q function, especially in the case of indistinguishable action-value differences between various actions of the same state. 
%To realize the duelling architecture, a duelling layer is placed before the output layer of the D3QN, separating the estimates of state-value and action-advantages into two different streams. Then, via aggregating the aforementioned two streams, the approximations of the state-action values can be delivered into the output layer. 
To further speed up and stabilize the learning process, $N_{ms}$-step learning and target network techniques are adopted for updating parameters of the online D3QN. Specifically, the online D3QN aims to minimize the following loss function
\begin{equation}
\mathcal{L}(\boldsymbol{\theta}_{D3})=
\left[	r_{t:t+N_{ms}} +\gamma^{N_{ms}}Q(\vec{q}_{u(t+N_{ms})},\vec{v}_{u}^{*}\vert\boldsymbol{\theta}^{-}_{D3})-Q(\vec{q}_{u(t)},\vec{v}_{u(t)}\vert\boldsymbol{\theta}_{D3})\right]^2, \label{loss_function_w_target_multistep_d3qn}
\end{equation} 
where $\boldsymbol{\theta}_{D3}$ is the parameter vector of the online D3QN, $\boldsymbol{\theta}_{D3}^{-}$ means the parameter vector of the target D3QN. The selected action $\vec{v}^{*}_{u}$ in (\ref{loss_function_w_target_multistep_d3qn}) is chosen from the online D3QN rather than the target D3QN, i.e., $\vec{v}^{*}_{u}=\underset{\vec{v}_{u}\in\mathcal{A}}{\arg\max}Q(\vec{q}_{u(t+N_{ms})},\vec{v}_{u}\vert\boldsymbol{\theta}_{D3})$, which completes the DDQN procedure. 

\textbf{Algorithm} \textbf{1} starts with network and hyper-parameter initializations, as shown in step 1. At the beginning of  each training episode, the UAV's initial location is randomly picked from the state space $\mathcal{S}$ (step 3). Then, the UAV chooses an action following the popular $\epsilon$-greedy action selection policy, which means that the UAV either selects a random action from the action space $\mathcal{A}$ with probability $\epsilon\in[0,1]$ or chooses the optimal action that maximizes the state-action approximation of the online D3QN with probability $1-\epsilon$. After the execution of the selected action, the environment will feed back the next state and the corresponding immediate reward (step 5). The experienced transition $exp_{n}$ will then be recorded by a sliding buffer, to prepare for the $N_{ms}$-step learning (step 17). When the sliding buffer is full, the latest $N_{ms}$-step experience can be generated and then delivered into the QiER buffer (step 18-step 24).
%, while being associated with order index that specifies the location of this transition in the QiER buffer. Besides, the corresponding qubit should be initialized as the eigenstate $\ket{0}$, to allocate the highest priority for this newly-recorded experience. Because the current $N_{ms}$-step transition has not been sampled for mini-batch training the online D3QN, the corresponding replay time should be reset to 0 (step 20)
Each training episode terminates when one of the following cases are encountered: reaching the destination, hitting the boundary, or exhausting the step threshold (step 26). When one episode is over, the exploration parameter $\epsilon$ will be annealed to encourage exploitation from exploration. For every fixed amount of training episodes, the target D3QN will be updated to the online counterpart (step 27). Once the QiER buffer is fully occupied, the mini-batch training for the online D3QN begins (step 6-step 16). With the mini-batch samples, the online D3QN is trained to minimize the mean counterpart of loss function (\ref{loss_function_w_target_multistep_d3qn}), via standard stochastic gradient descent approach (step 15).

\begin{algorithm}[htbp]
	\label{DRL_QiER_algorithm}
	\tiny
	\caption{The Proposed DRL-QiER Solution}
	\SetKwData{Ini}{\bf{Initialization:}}
	\Ini Initialize the online D3QN network $Q_{D3}(s,a\vert\boldsymbol{\theta}_{D3})$ and its target network $Q_{D3}(s,a\vert\boldsymbol{\theta}_{D3}^{-})$, with $\boldsymbol{\theta}_{D3}^{-}\leftarrow\boldsymbol{\theta}_{D3}$. Initialize the QiER buffer R with capacity $C$. Initialize the vector of replay time as $\vec{rt}=[rt_{1},rt_{1},\dots,rt_{C}]=\vec{0}$. Set the size of mini-batch as $N_{mb}$. Set the order index of R as $k = 1$. Set the flag indicating whether the QiER buffer is fully occupied or not as $LF=False$. Set the maximum TD error as $\delta_{\mathrm{max}} = 1$.\;\par
	\For{$te=[1,te_{\mathrm{max}}]$}{\par
		Set time step $n=0$.
		Randomly set the the UAV's initial location as $\vec{q}_{u}(n)\in\mathcal{S}$.
		Initialize a sliding buffer $\hat{\text{R}}$ with capacity $N_{ms}$.\;\par
		\Repeat{$\vec{q}_{u}(n)=\vec{q}_{u}(F)$ $\vert\vert$ $\vec{q}_{u}(n)\notin\mathcal{S}$ $\vert\vert$ $n = N_{\mathrm{max}}$}
		{	
			Select and execute action $a_{n}$, then observe the next state $\vec{q}_{u}(n+1)$ and  the immidiate reward $r_{n}=r_{n}[\vec{q}_{u}(n+1)]$\;
			\If{$LF==True$}{
				Perform quantum measurement on all stored experiences' qubits and get the vector of their replaying probabilities $[bp_{1},bp_{2},\dots,bp_{C}]$\;
				\For{$n_{mb}=[1,N_{mb}]$}{
					Sample a transition according to $[bp_{1},bp_{2},\dots,bp_{C}]$ and get its location index $d\in\{1,2,\dots,C\}$\;
					Reset the $d$-th qubit back to uniform quantum state $\ket{\Psi_{d}}=\ket{+}$\;
					Update the corresponding replay time $rt_{d}+=1$ and $rt_{\mathrm{max}}=\max(\vec{rt})$\;
					Calculate the sampled transition's absolute $N_{ms}$-step TD error $\vert\delta_{N_{ms}}\vert$ and update the maximum TD error $\delta_{\mathrm{max}} = \max(\delta_{\mathrm{max}}, \vert\delta_{N_{ms}}\vert)$\;
					Perform quantum preparation phase on the $d$-th qubit\;
					
				}
			Update the online D3QN network $Q_{D3}(s,a\vert\boldsymbol{\theta}_{D3})$ via gradient descent method using the mini-batch of sampled $N_{mb}$ transitions from R\;
			}		
		Get and record transition $exp_{n}=\{\vec{q}_{u}(n),a_{n},r_{n},\vec{q}_{u}(n+1)\}$ into $\hat{\text{R}}$\;
		\If{$n\geq N_{ms}$}{Generate the $N_{ms}$-step reward $r_{n-N_{ms}:n}$ from $\hat{\text{R}}$ and record $N_{ms}$-step experience $exp_{n-N_{ms}:n}=\{\vec{q}_{u}(n-N_{ms}),a_{n-N_{ms}},r_{n-N_{ms}:n},\vec{q}_{u}(n)\}$ into R with order index $k$\;	
			Perform quantum initialization phase on the $k$-th qubit as $\ket{\Psi_{k}} = \ket{0}$.
			Reset $rt_{k} = 0$ and let $k+=1$\;
			\If{$k>C$}{Set $LF=True$ and reset $k=1$\;}}
		Let $n+=1$\;}
		Update $\epsilon\leftarrow\epsilon \times dec_{\epsilon}$.
		Update the target D3QN $Q_{D3}(s,a\vert\boldsymbol{\theta}^{-}_{D3})$ every $\Upsilon_{D3}$ episodes, i.e., $\boldsymbol{\theta}^{-}_{D3}\leftarrow\boldsymbol{\theta}_{D3}$\;
	}
\end{algorithm}

\section{Numerical Results}
In this section, simulation results for the proposed DRL-QiER solution and the corresponding performance comparison against several baselines are performed. 
\subsection{Simulation Environment Setups}
\begin{figure}[htbp]
	\centering  
	\subfigcapskip = -.5cm
%	\vspace{-.5cm}
	\subfigure[\tiny The simulation environment]{
		\label{The simulation environment}
		\includegraphics[scale=0.26]{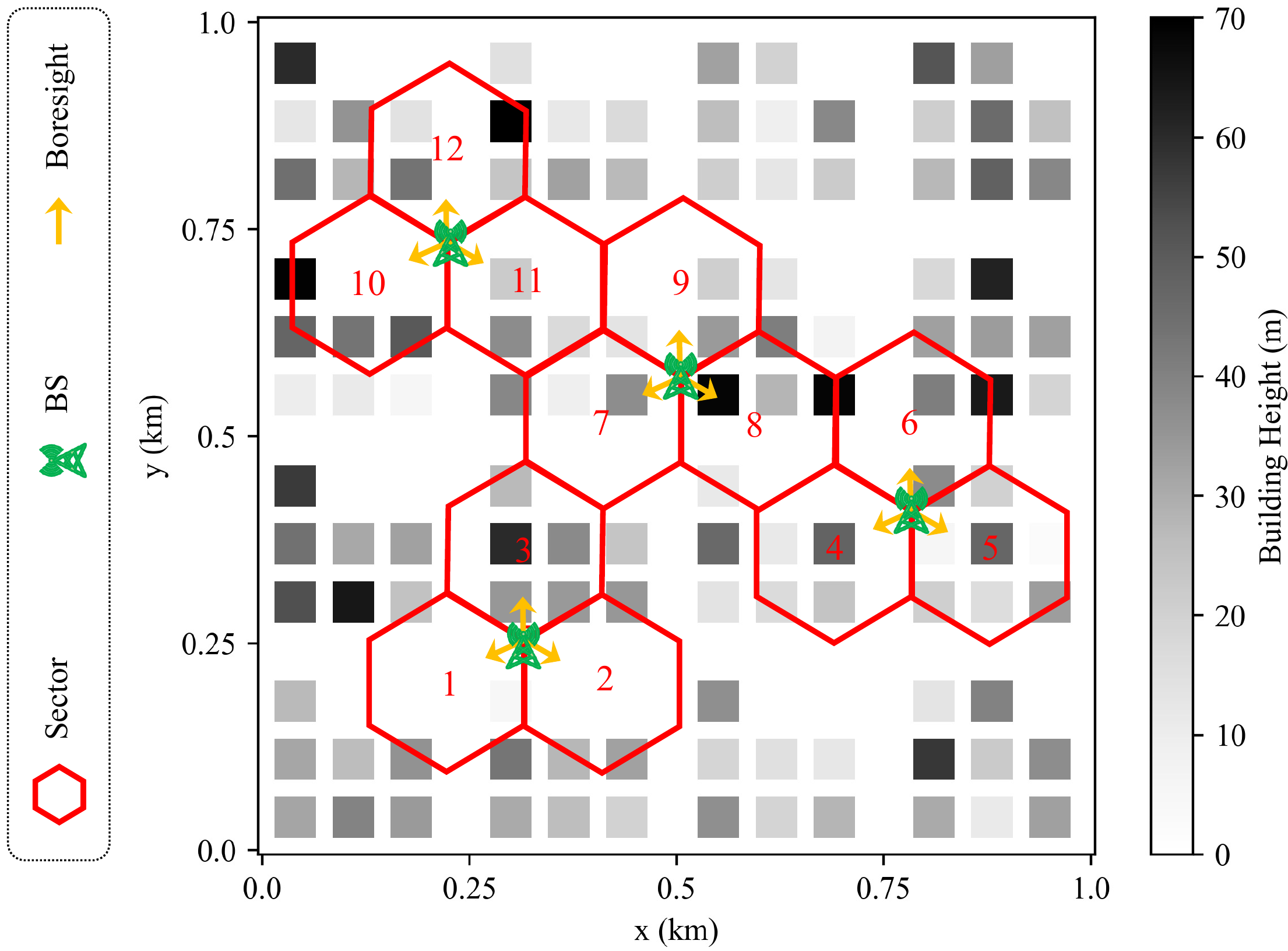}}\hspace{2cm}
	\subfigure[\tiny The corresponding TOP distribution]{
		\label{The corresponding TOP distribution}
		\includegraphics[scale=0.4]{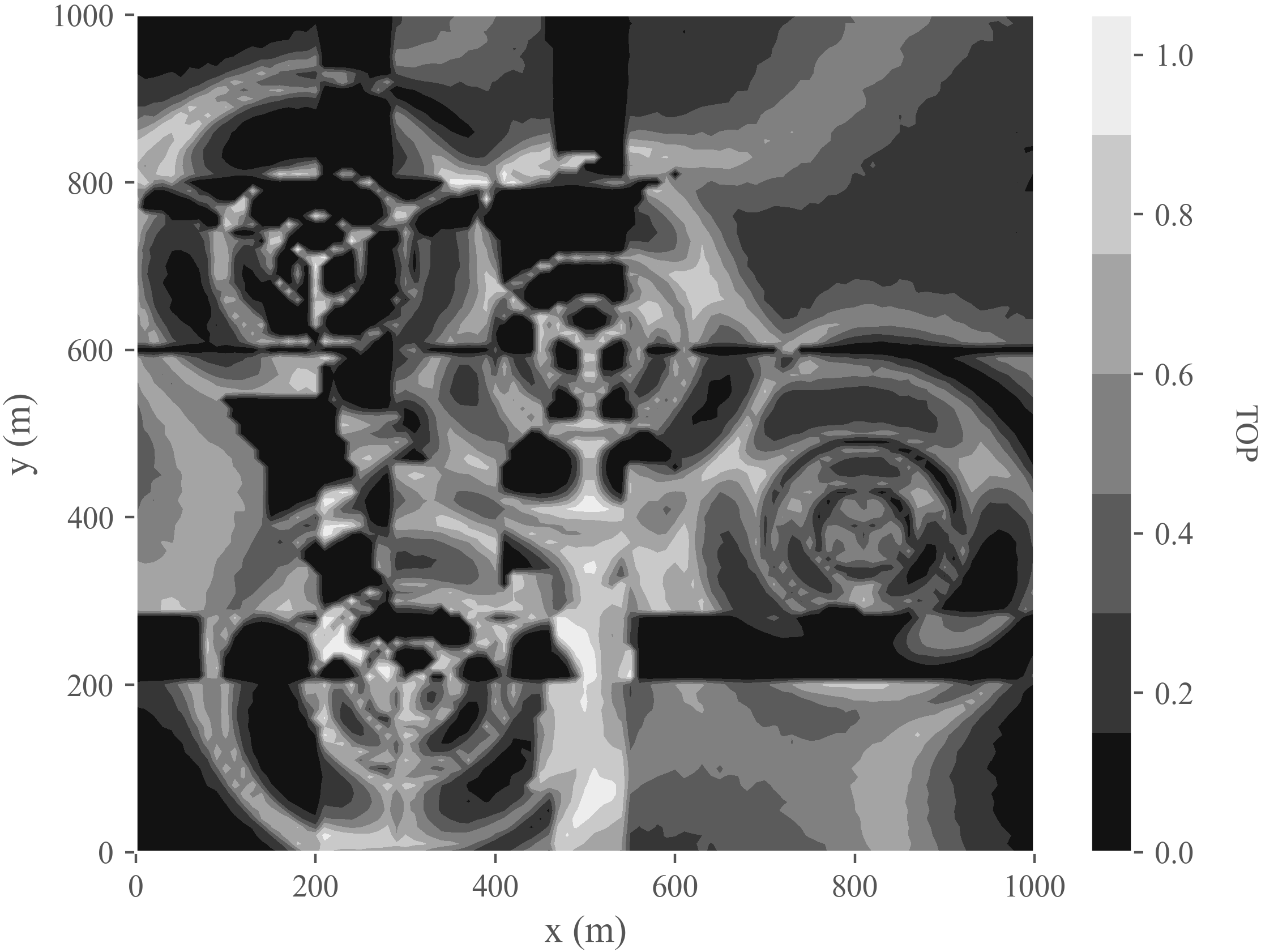}}\captionsetup{font={scriptsize}}
	\vspace{-.5cm}
	\caption{Simulation environment and the corresponding preview on TOP distribution}
%	\vspace{-1cm}
	\label{Simulation environment and the corresponding TOP distribution}
\end{figure}

For conducting the simulation, the UAV's exploration airspace is set as $\mathbb{A}:[0,1]\times[0,1]\times[0,0.1]$ km. Fig. \ref{The simulation environment} delivers the top view of $\mathbb{A}$, in which the locations of involved BSs and the direction of each ULA's boresight are specified. 
%The antenna height of each BS is chosen as 25 m, while the altitude of UAV is fixed as 100 m. 
%Each ULA contains $M=8$ directional array elements, with $\Theta_{3\text{dB}} = \Phi_{3\text{dB}} = 65\degree$, $d_{v} = \lambda/2$ where  
To generate building distribution within $\mathbb{A}$, one realization of statistical model suggested by the International Telecommunication Union (ITU) \cite{series2013propagation} is invoked, which is subject to the following three parameters: 1) $\hat{\alpha}$ indicates the ratio of region covered by buildings to the whole land; 2) $\hat{\beta}$ represents average amount of buildings; and 3) $\hat{\gamma}$ determines building heights' distribution (say, Rayleigh distribution with mean $\hat{\gamma}>0$). Besides, the small-scale fading component of G2A link is assumed to follow block Nakagami-$m$ channel model. 
%, which means that the channel remains static within one coherence interval and varies independently among different coherence intervals.
The common destination's location is fixed at $\vec{q}_{u} = (0.8,0.8,0.1)$ km, without loss of generality.
%For simplicity, the UAV is assumed to have $N_{fd}=4$ possible flying directions each time, i.e., forward, backward, left and right. 

Unless otherwise mentioned, the parameter setups regarding simulation environment are in line with Table \ref{Simulation_Settings}. With the generated local building distribution, antenna model and small-scale fading model, the corresponding TOP distribution over arbitrary UAV location within $\mathbb{A}$ can be previewed as Fig. \ref{The corresponding TOP distribution}. 
%In this paper, we set $\hat{\alpha}=0.3$, $\hat{\beta}=55$ buildings/km$^2$ and $\hat{\gamma}=25$ m. As a result, the total amount of buildings is $\hat{\beta}D^{2}=220$ and the expected size of each building is $\hat{\alpha}/\hat{\beta}\approx0.0055$ km$^2$. Besides, the maximum height of buildings is clipped to be under 70 m. 
\begin{table}[htbp] 
	\centering
	\captionsetup{font={scriptsize}}
	\tiny
	\vspace{-.5cm}
	\caption{Parameter Settings for Simulation Environment}
	\vspace{-.3cm}
	\label{Simulation_Settings}
	\begin{tabular}{p{5.4cm}|p{2.1cm}||p{5.5cm}|p{2cm}} 
		\toprule
		\textbf{Parameters} & \textbf{Values}  & \textbf{Parameters} & \textbf{Values}\\ 
		\midrule
		Amount of BSs $B$ & 4 & Amount of sectors $3B$ & 12\\
		
		Horizontal side-length of $\mathbb{A}$ $D$ & 1 km & Amount of each ULA's array elements $M$ & 8 \\
		
		Half-power beamwidth $\Theta_{3\text{dB}}/\Phi_{3\text{dB}}$ & $65\degree$/$65\degree$ & speed of light $c$ & $3\times 10^{8}$ m/s \\
		
		Carrier frequency $f_{c}$ & 2 GHz & Wave length $\lambda$ & 15 cm \\
		
		ULA's element spacing distance $d_{v}$ & 7.5 cm & ULA's electrically titled angle $\theta_{etilt}$ & 100$\degree$ \\
		
		Antenna height of BS & 25 m & Flying altitude of UAV & 100 m\\
		
		ITU building distribution parameter $\hat{\alpha}$ & 0.3 & ITU building distribution parameter $\hat{\beta}$  & 118   \\
		
		ITU building distribution parameter $\hat{\gamma}$ & 25 & total amount of buildings  $\hat{\beta}D^{2}$ & 118 \\
		
		expected size of each building $\hat{\alpha}/\hat{\beta}$ & 0.0025 km$^2$ & Maximum height of buildings & 70 m \\		
		Transmit power of each sector $P_{i}$ & 20 dBm & Nakagami shape factor $m$ for LoS/NLoS & 3/1\\	
		Transmission outage threshold $\Gamma_{th}$ & 0 dB  & Average power of AWGN $\sigma^2$ & -90 dBm\\		
		Duration of time slot $\Delta_{t}$ & 0.5 s  & Velocity of the UAV $V_{u}$ & 30 m/s\\
		Amount of signal Measurements $L$ & 1000  & Weight balancing the minimization $\tau$ & 50\\				
		\bottomrule
	\end{tabular}
\vspace{-.5cm}
\end{table}

\begin{comment}
For UAV at location $\vec{q}_{u} = (x_{u},y_{u},z_{u})$ and arbitrary BS $b\in\mathcal{B}$ at location $\vec{q}_{b} = (x_{b},y_{b},z_{b})$, the azimuth and elevation angles can be straightforwardly calculated as
\begin{equation}
\theta_{\text{G2A}} = 90\degree - \arctan(\frac{z_{u} - z_{b}}{\sqrt{\left(x_{u} - x_{b}\right)^{2} + \left(y_{u} - y_{b}\right)^{2}}})\in[0,90\degree),
\end{equation}
\begin{equation}
\phi_{\text{G2A}} = \arctan(\frac{y_{u} - y_{b}}{x_{u} - x_{b}}) +
\begin{cases}
0\degree,&x_{u}>x_{b}, y_{u}\geq y_{b}\\
180\degree,&x_{u}<x_{b}\\
360\degree,&x_{u}>x_{b}, y_{u}< y_{b}
\end{cases}
+
\begin{cases}
180\degree,&\text{Sector A}\\
60\degree,&\text{Sector B}\\
300\degree,&\text{Sector C}
\end{cases}
\end{equation}
\end{comment}

\subsection{Structure of DNNs and Hyper-parameter Settings for Learning Process}
The proposed DRL-QiER algorithm is implemented on Python 3.8 with TensorFlow 2.3.1 and Keras. Specifically, the DNNs of online D3QN agent are constructed with fully-connected feedforward ANNs. The shapes of the online D3QN's input and output layers are subject to the UAV's horizontal locations and the amount of possible flying directions, respectively. Between the input and output layers, there are 4 hidden layers, where the first 3 hidden layers contain 512, 256, 128 neurons, respectively. The last hidden layer plays the role as duelling layer consisting of $N_{fd} + 1$ neurons, where one neuron indicates the estimation of state-value and the other $N_{fd}$ neurons reflect action advantages. Then, the outputs of the duelling layer will be aggregated to generate the estimation of the $N_{fd}$ actions at the output layer. Besides, the optimizer minimizing the mean square error (MSE) for the DRL-QiER agent is Adam with fixed learning rate. The activation functions for each hidden layer and the ouput layer are Relu and Linear, respectively. Note that the target D3QN shares the same structure as its online counterpart.

The settings of hyper-parameter for learning process are stated in Table \ref{Learning_Settings}.
\begin{table}[htbp] 
	\tiny
	\captionsetup{font={scriptsize}}
	\centering
	\vspace{-.5cm}
	\caption{Hyper-parameter Settings for Learning Process} 
	\vspace{-.3cm}
	\label{Learning_Settings}
	\begin{tabular}{p{5.4cm}|p{2.1cm}||p{5.5cm}|p{2cm}} 
		\toprule
		\textbf{Parameters} & \textbf{Values}  & \textbf{Parameters} & \textbf{Values}\\ 
		\midrule
		Capacity of QiER buffer $C$ & 20000 & Size of mini-batch $N_{mb}$ & 128\\
		
		Initial $\epsilon$-greedy factor $\epsilon$ & 0.5 & Annealing speed $dec_{\epsilon}$& 0.554/episode \\
		
		Target D3QN update frequency $\Upsilon_{D3}$ & 5 & Length of sliding buffer $N_{ms}$ & 30 \\
		
		Positive special reward $r_{D}$ & 400  & Negative special reward $r_{ob}$  & -10000\\		
		Learning rate $\alpha_{lr}$ & Adam's default  & Discount factor $\gamma$  & 1\\		
		Maximum training episodes $te_{\text{max}}$ & 2000  & Step threshold $N_{\text{max}}$  & 400\\		
		\bottomrule
	\end{tabular}
\vspace{-1cm}
\end{table}

\subsection{Training of the DRL-QiER algorithm}
\begin{figure}[htbp]
	\centering  
	\subfigcapskip = -.5cm
	\vspace{-.3cm}
	\subfigure[\tiny Training return history]{
		\label{DRL-QiRL training returns}
		\includegraphics[scale=0.4]{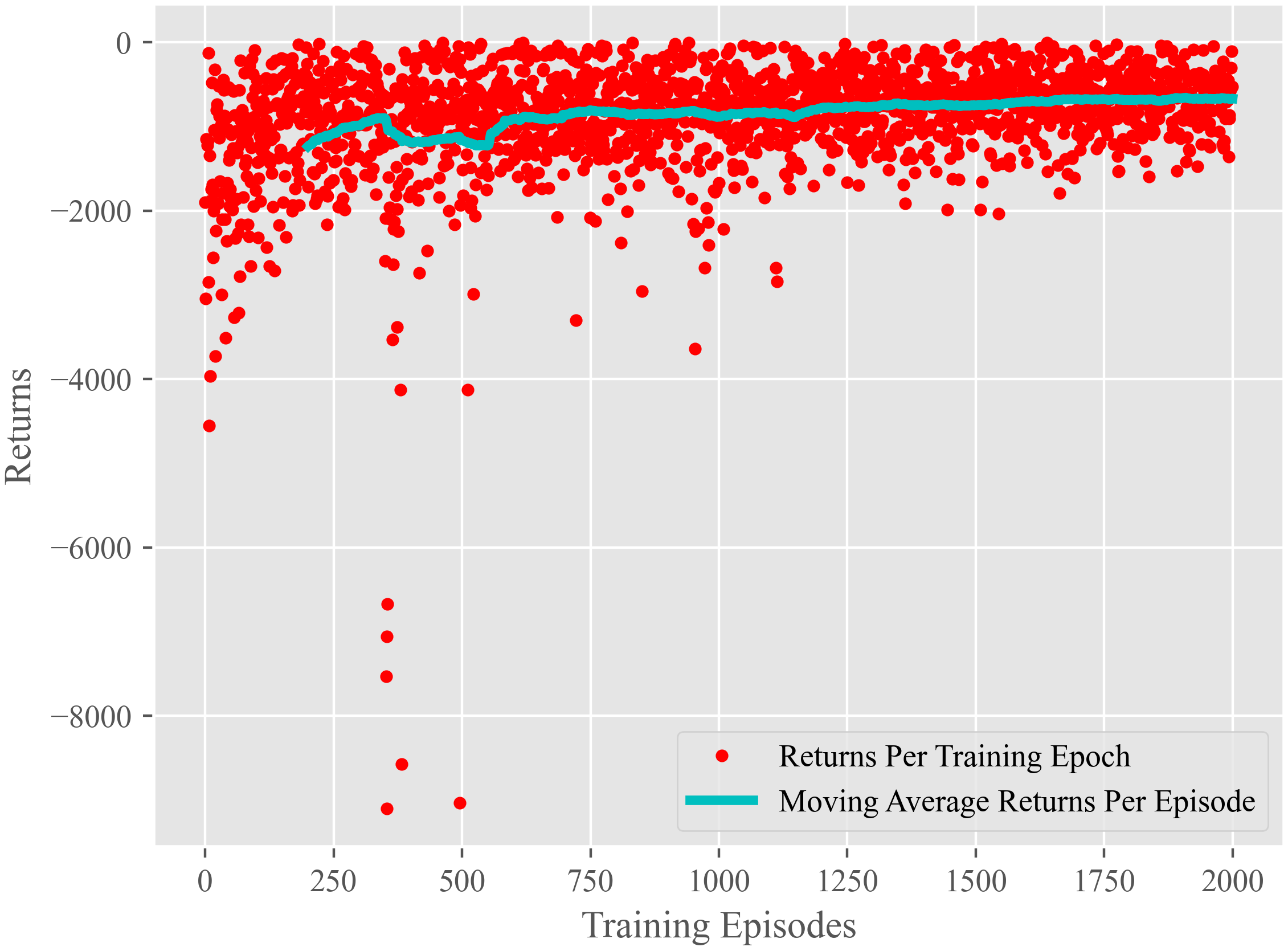}}\hspace{3cm}
	\subfigure[\tiny The corresponding designed trajectories]{
		\label{The corresponding designed trajectories}
		\includegraphics[scale=0.4]{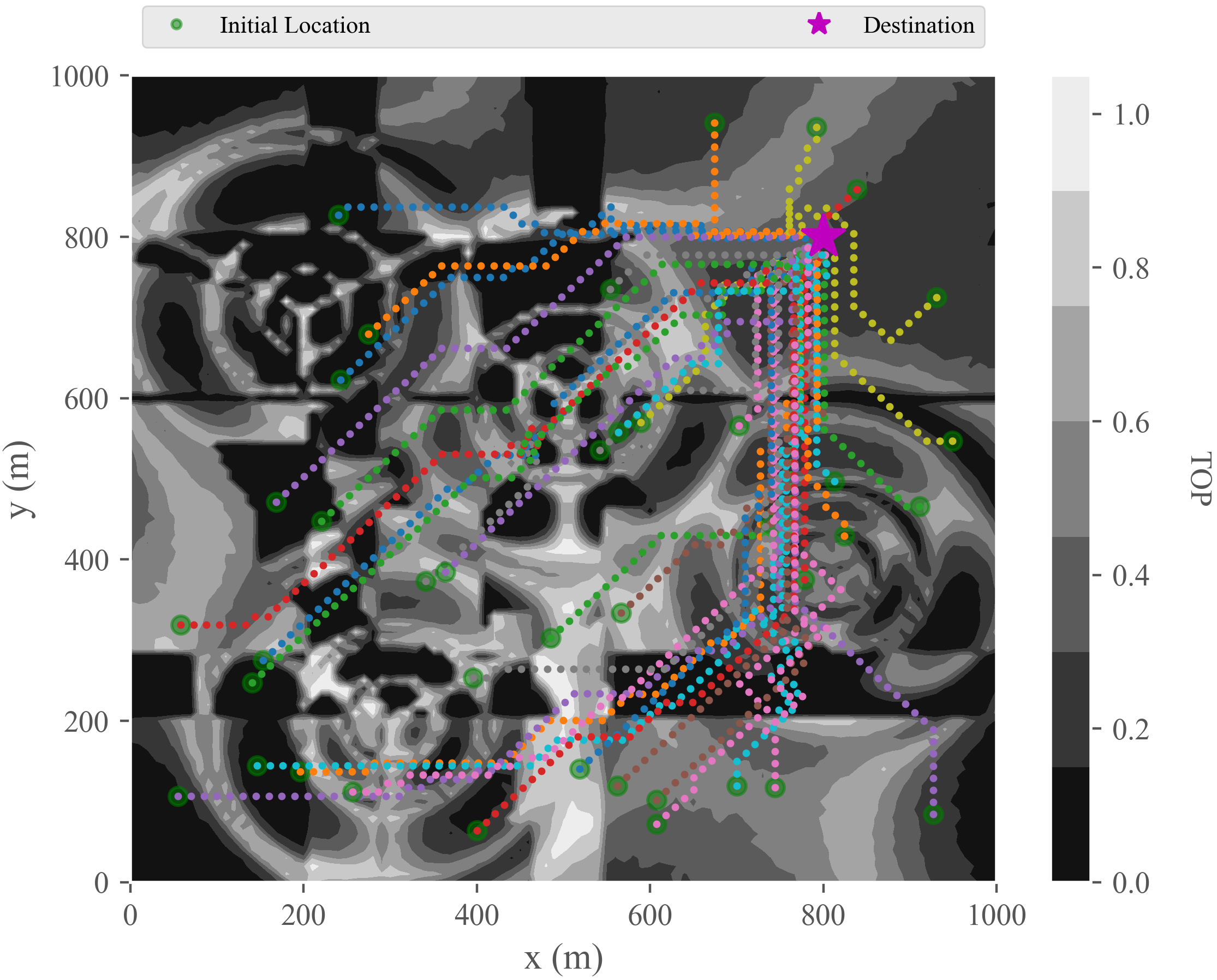}}\captionsetup{font={scriptsize}}
	\vspace{-.5cm}
	\caption{Training results of the proposed DRL-QiER solution}
	\vspace{-1cm}
	\label{Training results of the proposed DRL-QiER solution}
\end{figure}

Fig. \ref{DRL-QiRL training returns} and Fig. \ref{The corresponding designed trajectories} depict the return history and designed trajectories of the proposed DRL-QiER solution, respectively. Note that the moving average return for each training episode is calculated via a moving window with length of 200 episodes, while the corresponding designed trajectories are picked with spatially-separated initial locations in the late training stage (in the range of episodes 1900-2000), for the sake of neat and sufficient demonstration. From Fig. \ref{DRL-QiRL training returns}, it is straightforward to conclude that the moving average returns steadily converge to the maximum alongside the training process, although some fluctuations are experienced, which is a typical phenomenon in DRL field. Besides, from Fig. \ref{The corresponding designed trajectories}, it is observed that the proposed DRL-QiER solution can direct the UAV from various initial locations to the common destination, with designed trajectories adaptive to the TOP distribution. Regions with higher TOP are avoided while keeping the UAV being directed to reach the common destination with possibly fewer moving steps (equivalently, as short flying time cost as possible). For instance, even the near-to-zero but extremely narrow TOP slots around $(0.4, 0.78, 0.1)$ km and $(0.6, 0.79, 0.1)$ km can be recognized. On the contrary, higher TOP regions in the range of $(0.4-0.6, 0-0.5, 0.1)$ km are bypassed as much as possible. Another good example is the trajectory starting from location around $(0.95, 0.09, 0.1)$ km, where the "V" shape around $(0.95, 0.2, 0.1)$ km perfectly demonstrates the effectiveness of the proposed DRL-QiER solution, in which the higher TOP fields are avoided. Note that larger weight factor $\tau$ will generally lead the designed path to experience lower TOP regions, but inevitably enlarging the time cost (say, longer and more tortuous trajectory) reaching the common destination. This is the reason why weight factor $\tau$ is invoked to balance the proposed minimization problem (P1).
\vspace{-.5cm}
\subsection{Performance Comparison}
Four DRL-aided baselines are considered for performance comparison, listed as follows.
\begin{itemize}
	\item \textit{DRL-ER}: The D3QN is trained via mini-batch sampling from standard ER buffer with uniform sample-picking strategy.
%	, which means that the transitions are picked randomly from the ER buffer to accomplish the mini-batch sampling process.
	\item \textit{DRL-PER}: The D3QN is trained via stochastic mini-batch sampling from the PER buffer with proportional prioritization method, in line with \cite{schaul2015prioritized}. In this approach, the priority of each recorded transition $x$ is measured by its corresponding absolute TD error $\vert\delta(x)\vert$. Then, the probability of picking a transition from the PER buffer follows $p(x)=(\vert\delta(x)\vert + \xi)^{\alpha_{\text{PER}}}/\sum_{x'}(\vert\delta(x')\vert + \xi)^{\alpha_{\text{PER}}}$, where a small positive constant $\xi$ is used to avoid zero-priority case and $\alpha_{\text{PER}}$ determines how much prioritization is applied, with $\alpha_{\text{PER}}=0$ corresponding to the special case that is equivalent to DRL-ER baseline. To correct the bias caused by priority-based sampling, normalized importance-sampling (IS) weight $W(x)=(C\times p(x))^{-\beta_{\text{PER}}}/\max_{x'}W(x')$ is calculated to scale the updating of DNNs, where $C$ is the capacity of the PER buffer and $\beta_{\text{PER}}$ reflects the amount of IS correction. The parameter $\beta_{\text{PER}}$ should be incremented from a relatively small positive constant to 1 over the training process because a full-step update is more important when the algorithm begins to converge. 
	\item \textit{DCRL}: The DCRL training paradigm aims to offer better mini-batch sampling efficiency, according to the complexities of recorded experiences. Specifically, the complexity of each transition is determined by self-paced priority and coverage penalty, where self-paced priority maps TD error into the difficulty of current curriculum and coverage penalty uses replay times of transitions to enhance sampling diversity. For detailed implementation of DCRL, please refer to \cite{ren2018self}.
	\item \textit{SNARM}: The framework SNARM invokes an extra DNN termed as radio map to help improve the overall learning efficiency. The signal measurements alongside the UAV's trajectory are utilized to train not only the online D3QN but also the radio map. The radio map enables it to generate simulated trajectories and thus reduces actual trials. Based on standard Dyna architecture, one D3QN update with the actual experiences follows several extra updates with the simulated transitions. Therefore, the SNARM approach is promised to help achieve better learning performance while reducing the cost of data acquisition from actual experiences. For more details of SNARM, please refer to \cite{zeng2021simultaneous}.
\end{itemize}

For fair comparison, the structures of online and target D3QNs for all baselines are the same as those of the proposed DRL-QiER solution, while the hyper-parameter settings of these baselines are in line with Table \ref{Learning_Settings}. Besides, the construction of radio map's DNN and the corresponding hyper-parameter settings of baseline SNARM are in accordance to \cite{zeng2021simultaneous}, while the complexity index function, the curriculum evaluation function, the self-paced prioritized function, the coverage penalty function and the corresponding DCRL hyper-parameter settings are in line with \cite{ren2018self}. Furthermore, the additional hyper-parameters regarding PER in DRL-PER baseline are set as $\alpha_{\text{PER}}=1$, $\xi=0.01$ and $\beta_{\text{PER}}=0.4$. All the baselines are altered to involve multi-step learning and start training after their replay buffers are fully exploited. Nevertheless, all the baselines share the same randomly-generated initial UAV locations with the proposed DRL-QiER solution, for each training episode.

\begin{figure}[htbp]
	\centering  
	\subfigcapskip = -.5cm
	\vspace{-.2cm}
	\subfigure[\tiny Comparison on moving average returns]{
		\label{Performance comparison on moving average returns}
		\includegraphics[scale=0.4]{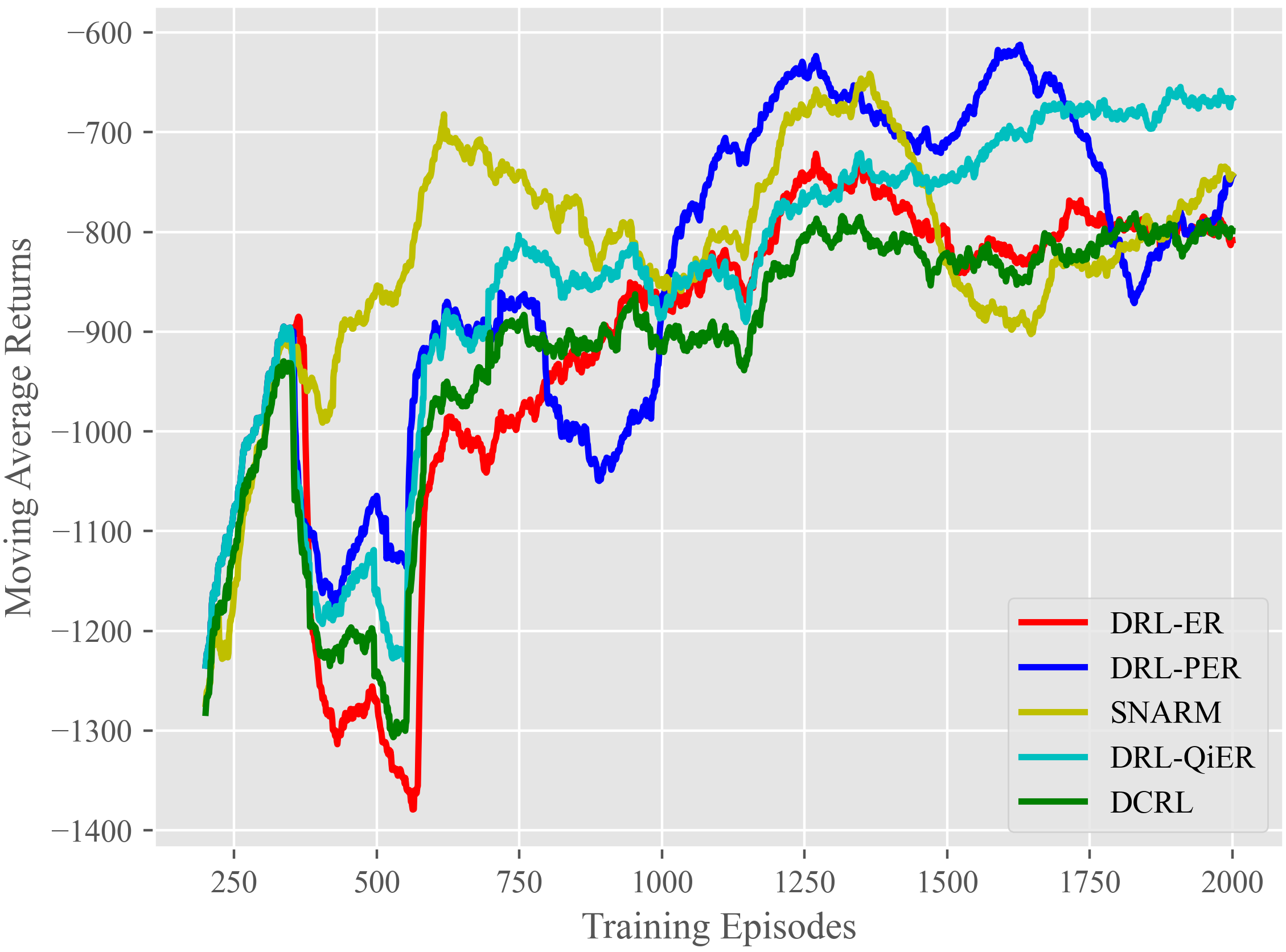}}\hspace{3cm}
%	\subfigure[\tiny Designed trajectories at early training stage]{
%		\label{Designed trajectories at early training stage}
%		\includegraphics[scale=0.4]{Performance_Comparison_Trajectories_early_stage.png}}%\captionsetup{font={scriptsize}}
%\\
%	\centering  
%	%\subfigcapskip=-.6cm
%	\subfigure[\tiny Designed trajectories at middle training stage]{
%		\label{Designed trajectories at middle training stage}
%		\includegraphics[scale=0.4]{Performance_Comparison_Trajectories_middle_stage.png}}\hspace{2cm}
	\subfigure[\tiny Designed trajectories of trained agents]{
		\label{Designed trajectories of trained agents}
		\includegraphics[scale=0.4]{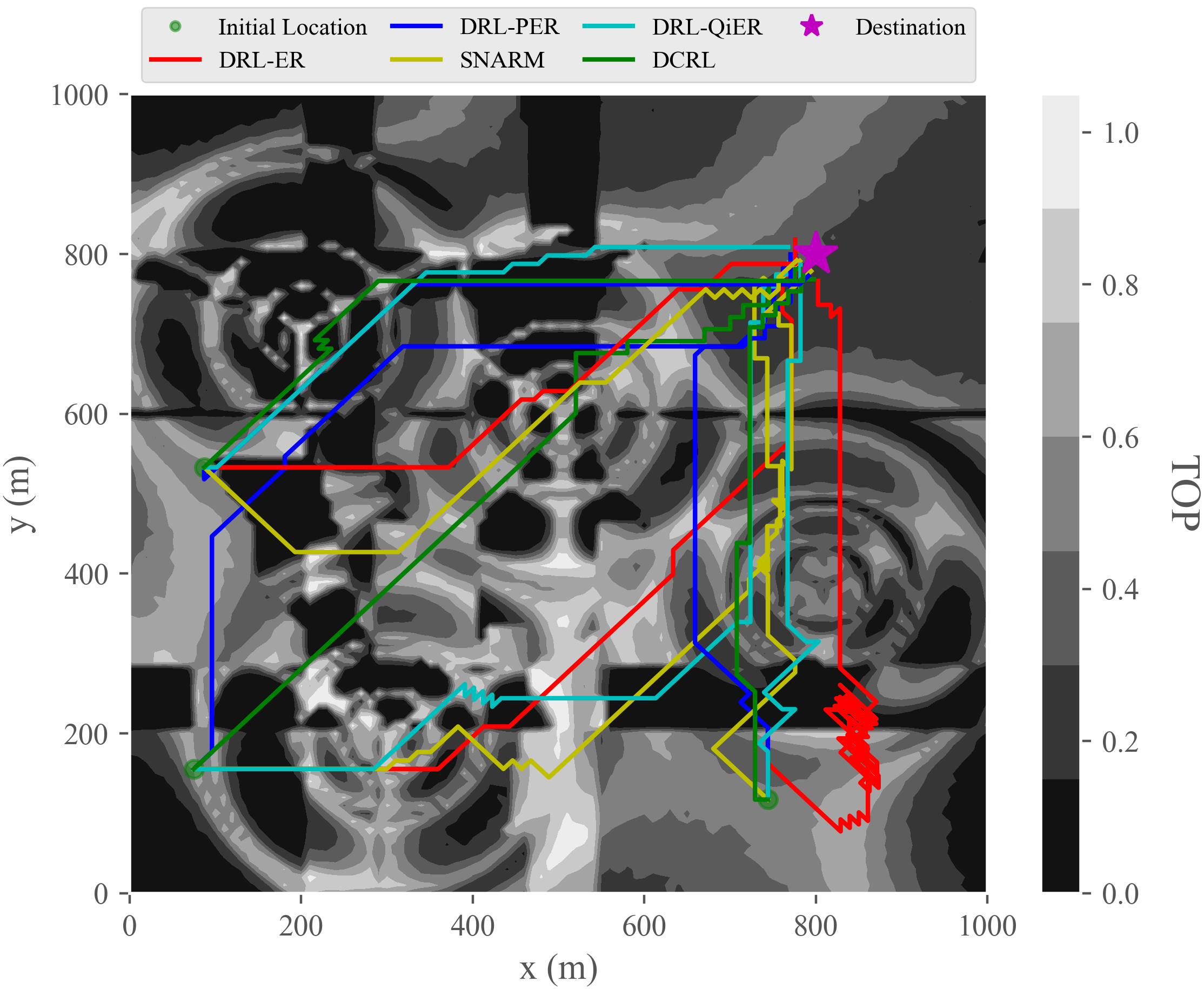}}\captionsetup{font={scriptsize}}
	\vspace{-.5cm}
	\caption{Performance comparison on moving average returns and designed trajectories}
	\vspace{-0cm}
	\label{Performance comparison on moving average returns and designed trajectories}
\end{figure}

Fig. \ref{Performance comparison on moving average returns} delivers the performance comparison on moving average returns of the proposed DRL-QiER solution and considered baselines, versus training episodes. From this subfigure, it is easy to find that SNARM approach can offer satisfactory learning performance, thanks to the simulated trajectories enabled by the extra DNN (i.e., the radio map). Especially, in the range of training episode from 400 to 1000, despite that the radio map is getting well trained as the training process going. Besides, DRL-PER, DRL-QiER and DCRL approaches can achieve better moving average returns than DRL-ER method, in the early training stage (e.g., episodes 500-750). The reason is that DRL-ER solution samples transitions uniformly without considering their priorities, which leads transitions with higher importance to have less opportunities for training the online D3QN. However, DRL-PER method experiences server fluctuations than DRL-QiER and DCRL (e.g., episodes 1250-2000), which is because DRL-PER does not take transitions' replay time into account and thus some transitions are sampled with undesired high frequency while their absolute TD errors remain relatively large. The proposed DRL-QiER solution showcases more steady learning ability, with less amplification of fluctuation and overall raising trend, thanks to the QiER technique which balances sampling priority and diversity in a better manner. Although SNARM and DCRL approaches can offer satisfactory learning performances, their respective shortcomings are: 1) SNARM framework needs to train an extra DNN, which thus introduces heavy training burden, and 2) it is difficult to set up feasible complexity index function, curriculum evaluation function, self-paced prioritized function, coverage penalty function and the corresponding DCRL hyper-parameters, which limits the robustness of DCRL solution. The proposed DRL-QiER method requires less hyper-parameters tuning and contains no extra DNN, and therefore is easier and more robust for implementation.
To deliver more insights, Fig. \ref{Designed trajectories of trained agents} depicts the comparison on designed trajectories of the implemented algorithms, over three representative starting locations chosen from episodes 1910-2000. It is straightforward to observe that the proposed DRL-QiER and the considered baselines direct the UAV to hit the common destination with different trajectories.  

\begin{figure}[htbp]
	\centering  
	\subfigcapskip = -.5cm
	\vspace{-.2cm}
	\subfigure[\tiny Comparison on average time cost]{
		\label{Average time durations and EOD over various episode slots}
		\includegraphics[scale=0.4]{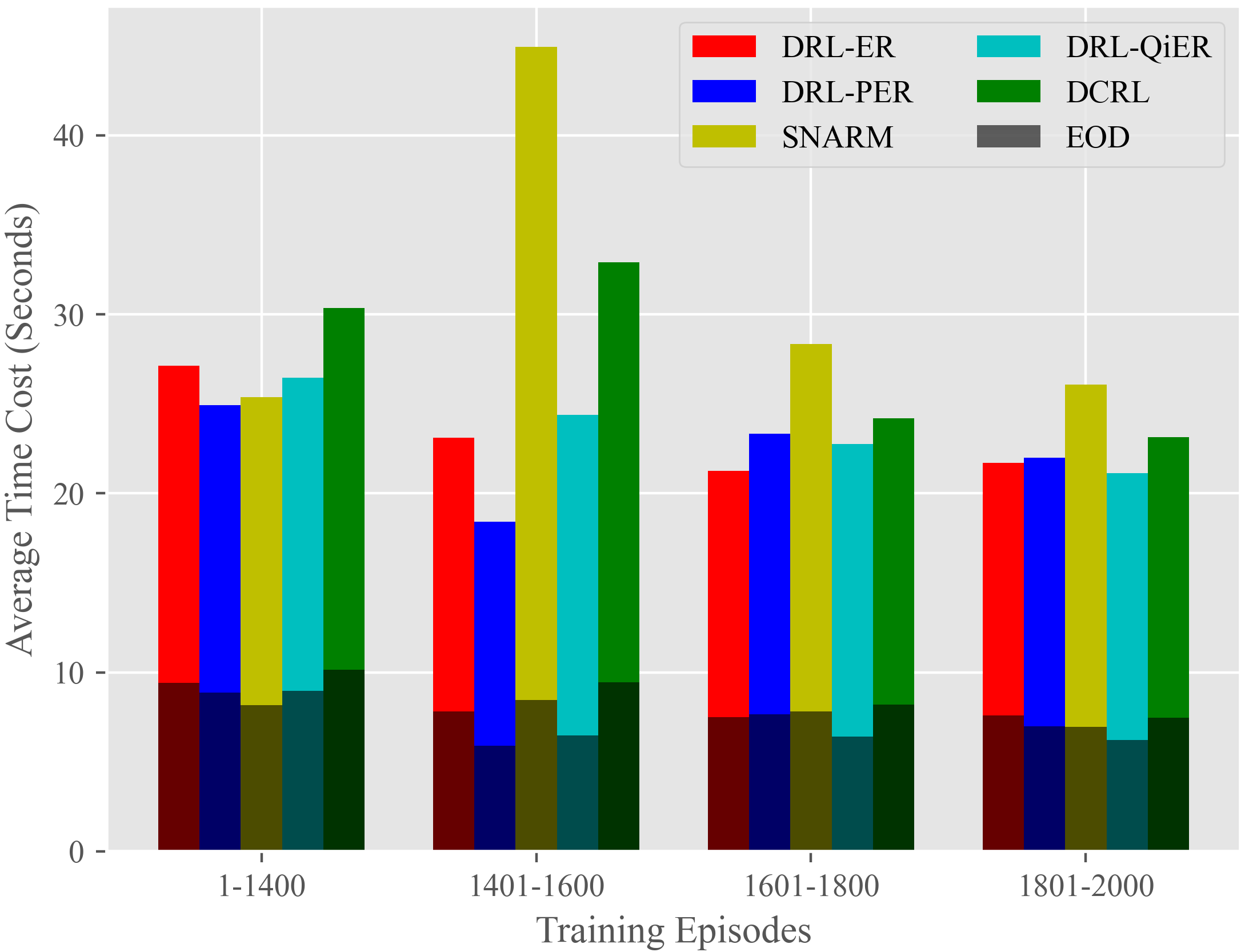}}\hspace{3cm}
	\subfigure[\tiny Comparison on average duration]{
		\label{Comparison on average duration over the last 200 episodes}
		\includegraphics[scale=0.37]{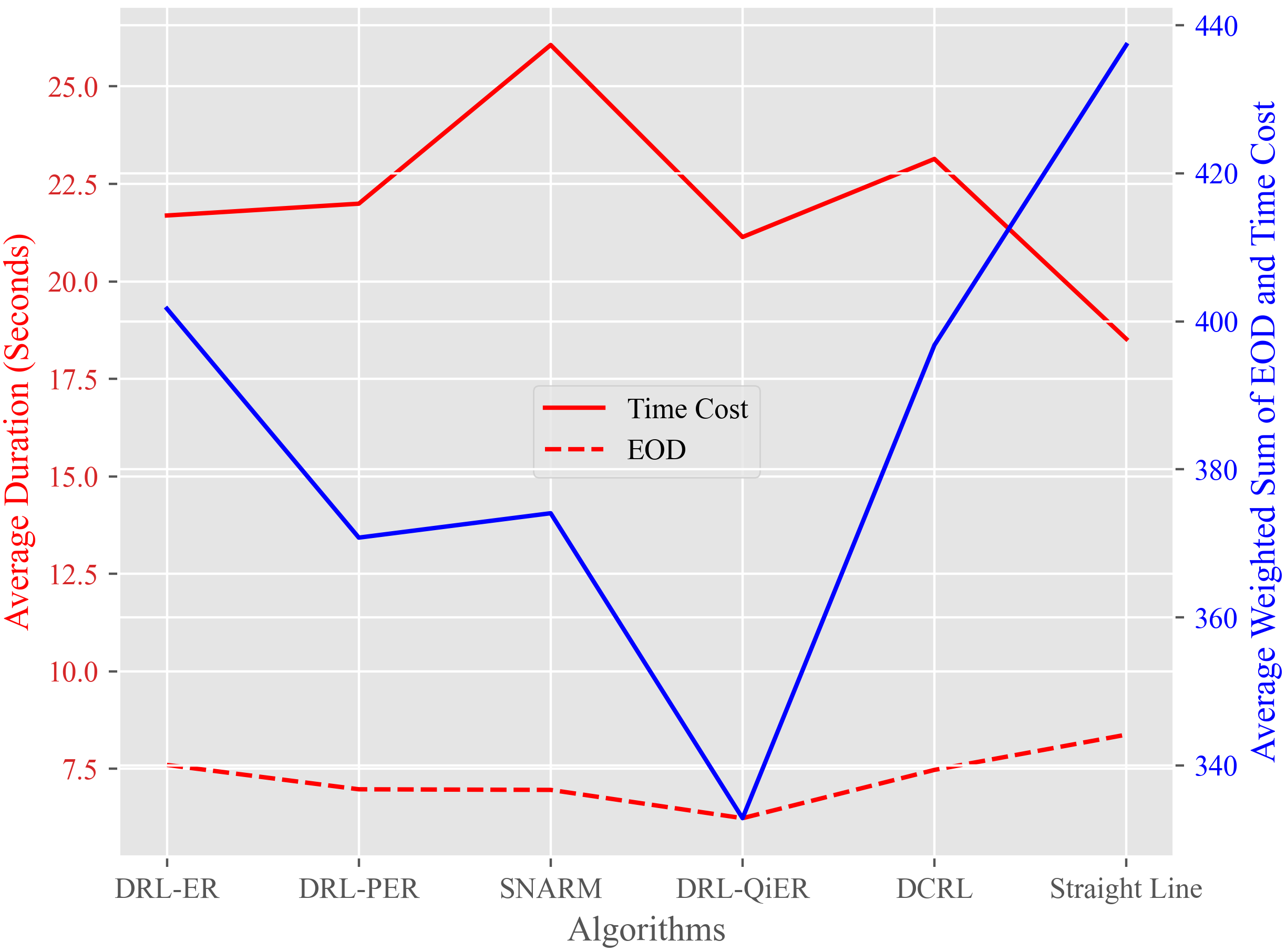}}\captionsetup{font={scriptsize}}
	\vspace{-.5cm}
	\caption{Performance comparison on average time costs and EOD}
	\vspace{-.5cm}
	\label{Performance comparison on average time cost and EOD}
\end{figure}

Fig. \ref{Average time durations and EOD over various episode slots} demonstrates comparison on average time cost of designed trajectories and the corresponding EOD for the considered algorithms, over four episode slots 1-1400, 1401-1600, 1601-1800 and 1801-2000. From this figure, one can find that the proposed DRL-QiER solution can help achieve both lower average EOD and average time cost, within each episode slot. Especially, in the late training state (e.g., episode slot 1800-2000), the proposed DRL-QiER method outperforms other baselines, in terms of both average EOD and average time cost. Furthermore, Fig. \ref{Comparison on average duration over the last 200 episodes} illustrates comparison on average duration and average weighted sum of EOD and time cost over the last 200 training episodes, for all the DRL-aided approaches and non-learning-based strategy termed as straight line. From this figure, it is easy to find that while the straight line solution offers the cheapest average time cost, it leads the UAV to suffer the highest average EOD, which is extremely non-preferable and thus unveil the benefits provided by DRL-aided approaches. On the contrary, the proposed DRL-QiER solution can not only help the UAV experience the lowest average EOD, compared to both other DRL-aided approaches and the straight line strategy, but also direct the UAV to reach the common destination with the cheapest average time cost, against other DRL-aided solutions. 
\vspace{-.5cm}
\section{Conclution}
In this work, an intelligent navigation task for cellular-connected UAV networks was investigated, aiming at minimizing the weighted sum of time cost and expected outage duration alongside UAVs' flying trajectories towards the common destination with randomly-generated initial UAV locations. To navigate the UAV, a DRL-QiER solution was proposed, in which the innovative QiER technique helps the DRL agent hit a better learning efficiency. Simulation results validated the effectiveness of the proposed DRL-QiER solution, while performance comparison against both several DRL-aided baselines and straight line strategy showcased DRL-QiER method's superiority. Moreover, the proposed QiER framework can be potentially extended into other existing DRL frameworks that are dependent on ER technique, e.g., deep deterministic policy gradient (DDPG), soft actor-critic (SAC) and Rainbow.
\bibliographystyle{IEEEtran}
\bibliography{reference-icc}

\end{document}